\documentclass{article}
\usepackage{amsfonts}
\usepackage{amssymb, amsmath}

\newtheorem{theorem}{Theorem}

\newtheorem{proposition}{Proposition}

\numberwithin{equation}{section}
\newenvironment{proof}[1][Proof]{\textbf{#1.} }{\ \rule{0.5em}{0.5em}}

\def\<{\hskip-2pt}
\def\({\hskip1pt(}
\def\){)\hskip1pt}
\def\>{\hskip1pt}

\def\?{\hskip-1.2pt,\hskip-0.6pt}

\input{tcilatex}

\begin{document}

\title{A Dynamical Theory of Quantum Measurement and Spontaneous Localization%
}
\author{Viacheslav~P.~Belavkin \\
Moscow Institute of Electronics and Mathematics,\\
B. Vuzovski\u{\i}\ per. 3/12, Moscow 109028, Russia}
\date{Received November 9, 1993\\
Published in \textit{Russian Journal of Math. Physics} \textbf{3} (1) 3--24
(1995) }
\maketitle

\begin{abstract}
We develop a rigorous treatment of discontinuous stochastic unitary
evolution for a system of quantum particles that interacts singularly with
quantum \textquotedblleft bubbles" at random instants of time. This model of
a "cloud chamber" allows to watch and follow with a quantum particle
trajectory like in cloud chamber by sequential unsharp localization of
spontaneous scatterings of the bubbles. Thus, the continuous reduction and
spontaneous localization theory is obtained as the result of quantum
filtering theory, i.e., a theory describing the conditioning of the a priori
quantum state by the measurement data. We show that in the case of
indistinguishable particles the a posteriori dynamics is mixing, giving rise
to an irreversible Boltzmann-type reduction equation. The latter coincides
with the nonstochastic Schr\"{o}dinger equation only in the mean field
approximation, whereas the central limit yields Gaussian mixing fluctuations
described by stochastic reduction equations of diffusive type.
\end{abstract}


\footnotetext{%
On leave of absence from Mathematics Department of Nottingham University,
NG7 2RD, UK.}

\section{Introduction}

The quantum measurement theory based on the ordinary von Neumann reduction
postulate applies neither to instantaneous observations with continuous
spectra nor to continual (continuous in time) measurements. Although such
phenomena can be described in the more general framework of Ludwig's
Davies-Lewies operational approach \cite{1}--\cite{5}, there is a particular
interest in describing quantum measurements by concrete Hamiltonian models
from which the operational description can be derived by an averaging
procedure. Perhaps the first model of such kind for instantaneous unsharp
measurement of particle localization was given by von Neumann \cite{6}. He
considered the singular interaction Hamiltonian 
\begin{equation}
h_{x}(t)=x\delta (t)\frac{\hbar }{\mathrm{i}}\frac{\mathrm{d}}{\mathrm{d}y}%
,\quad \delta (t)=%
\begin{cases}
\infty , & t=0, \\ 
0, & t\neq 0%
\end{cases}
\label{0.1}
\end{equation}%
with Dirac $\delta $-function, producing the translation operator 
\begin{equation}
s_{x}=\exp \Bigg\{-\frac{\mathrm{i}}{\hbar }\int_{-\infty }^{\infty
}h_{x}(t)\,\mathrm{d}t\Bigg\}=e^{-x\mathrm{d}/\mathrm{d}y}  \label{0.2}
\end{equation}%
at time $t=0$ of the measurement. Here $x$ is the position of the particle
and $y$ is the pointer position given by the $q$-coordinate of a quantum
meter. The particle scattering operator $S=\{s_{x}\}$ applied to the
(generalized) position eigen-vectors $|x\rangle $ as $S|x\rangle =|x\rangle
s_{x}$ does not affect the position $x$ of the particle but changes the
meter coordinate $q$ to $y=s_{x}^{\dagger }qs_{x}=x+q$. This implies that
the initial wave function $\psi _{0}(x,y)$ of the system \textquotedblleft
particle plus meter" is transformed into 
\begin{equation}
\psi (x,y)=s_{x}\psi _{0}(x,y)=\psi _{0}(x,y-x).  \label{0.3}
\end{equation}

If in the initial state the particle and the meter were not coherent, $%
\psi_0(x,y) = \eta(x) f_0(y)$, and the wave function $f_0$ of the meter was
fixed, then one can obtain the unitary transformation $S:\psi_0\mapsto
s_\bullet\psi_0$ via a family $\{F(y)\}$ of reduction transformations $F(y)
: \eta\mapsto f_\bullet(y)\eta$ for the particle vector-states $\eta$.
Specifically, Eq.~(\ref{0.3}) can be defined as 
\begin{equation}
\psi(x,y) = f_0(y-x)\eta(x)\equiv f_x(y)\eta(x).  \label{0.4}
\end{equation}
The linear nonunitary operators $F(y)=s_\bullet f_0(y)$ act on $|x\rangle$
as the multiplication $F(y)|x\rangle=|x\rangle f_x(y)$ by $%
f_x(y)=f_0(y-x)=s_xf_0(y)$ and would give a sharp localization of any
particle wave function $\eta(x)$ at the point $x=y$ of the pointer position
provided that the wave function $f_0(y)$ could be initially localized at $%
y=0 $. But there are no such sharply localized quantum states for the
continuous pointer, and the best that one can do is to take a wave packet $%
f_0(y)$, say, of the Gaussian form 
\begin{equation}
f_0(y) = \exp\bigg\{-\frac{\pi}{2}y^2\bigg\}.  \label{0.5}
\end{equation}
This results in the unsharp localization 
\begin{equation}
\psi(x,y) = \exp\biggl\{-\frac\pi2\,(x-y)^2\biggr\}\eta(x)  \label{0.6}
\end{equation}
of any particle wave function $\eta(x)$ about the observed value $y$ of the
pointer, normalized to the probability density 
\begin{equation}
{\mathrm{p}}_0(y) = \int|f_0(y-x)|^2 |\eta(x)|^2 \,\, x\,.  \label{0.7}
\end{equation}
For commuting operators $x$ and $y$, this is equivalent to the classical
measurement model $y = x+q$ for unsharp measurement of an unknown signal $x$
via the sharp measurement of the signal plus Gaussian noise $q$ with given
probability density $|f_0(q)|^2$.

These arguments illustrate how to interpret the reduction model involving
the continuous spectrum of a quantum measurement as a Hamiltonian
interaction model with nondemolition observation for a quantum object using
the measurement of the pointer coordinate of the quantum meter. Since von
Neumann introduced this approach, it was used by numerous other authors \cite%
{7, 8} for the derivation of a generalized reduction $\eta\mapsto F(y)\eta$
that would replace the von Neumann postulate $\eta\mapsto E(y)\eta$ given by
orthoprojections $\{E(y)\}$. As extended to nondemolition observations
continual in time \cite{9}--\cite{15}, this approach consists in using the
quantum filtering method for the derivation of nonunitary stochastic wave
equations describing the quantum dynamics under the observation. Since a
particular type of such equations has been taken as a postulate in the
phenomenological theory of continuous reduction and spontaneous localization 
\cite{16}--\cite{20}, the question arises whether it is possible to obtain
this equation from an appropriate Schr\"odinger equation. Here we shall show
how this can be done by second quantization of the interaction Hamiltonian
considered by von Neumann, obtaining a stochastic model of continual
nondemolition observation for the position of a quantum particle by counting
some other quanta. But first we show that even the projection postulate can
be derived in the framework of this approach with the suggested Hamiltonian
interaction and a proper nondemolition observation.

\section{Hamiltonian Reduction Model}

Let $\mathcal{H}$ be a Hilbert space, called the state space of a particle,
and let $\mathbf{R}=\{R^{\alpha }:\alpha =1,\ldots ,d\}$ be selfadjoint
operators in $\mathcal{H}$ with either integer or continuous spectrum $%
\mathbb{Z}^{d}$ or $\mathbb{R}^{d}$, $d\in \{1,2,3\}$. Let $\kappa >0$ be a
scaling parameter. If the operators commute, one can regard the scaled
vector operator $\kappa \mathbf{R}$ as the position $\mathbf{x}$ of the
particle in $\mathbb{R}^{d}$ or in the $d$-dimensional lattice $\kappa 
\mathbb{Z}^{d}$ if it is quantized. In the $\mathbf{x}$-representation, the
operator acts as the multiplication $\kappa R^{\alpha }|\mathbf{x}\rangle =|%
\mathbf{x}\rangle \lambda (x^{\alpha })$ by $\lambda (x^{\alpha })=\kappa
\lfloor x^{\alpha }/\kappa \rfloor $, where $\lfloor \mathbf{x}\rfloor \in 
\mathbb{Z}^{d}$ denotes the integer part of the vector $\mathbf{x}%
=\{x^{\alpha }:\alpha =1,\ldots ,d\}$.

A quantum meter with continuous ($\Lambda =\mathbb{R}^{d}$) or lattice ($%
\Lambda =\varepsilon \mathbb{Z}^{d}$) pointer scale is described by the
Hilbert space $L^{2}(\Lambda )$ of complex-valued functions $f:\Lambda
\rightarrow \mathbf{C}$ square integrable in the sense that $\Vert f\Vert
^{2}=\int |f(\mathbf{y})|^{2}\mathrm{d}\lambda <\infty $, where 
\begin{equation*}
\int f(\mathbf{y})\,\mathrm{d}\lambda =%
\begin{cases}
\sum_{y\in \Lambda }f(\mathbf{y})\varepsilon ^{d}\,, & \text{if }\Lambda
=\varepsilon \mathbb{Z}^{d}\,,\quad \mathrm{d}\lambda =\varepsilon ^{d} \\ 
\int_{\mathbf{y}\in \Lambda }f(\mathbf{y})\mathrm{d}\mathbf{y}\,, & \text{if 
}\Lambda =\mathbb{R}^{d}\,,\quad \ \mathrm{d}\lambda =\mathrm{d}y\,.%
\end{cases}%
.
\end{equation*}

Consider a moving particle with Hamiltonian $H$ in $\mathcal{H}$. Its
singular evolution corresponding to the position measurement at time $t=0$
is described in the product space $\mathcal{H}_{1}=\mathcal{H}\otimes
L^{2}(\Lambda )$ by the time-dependent Hamiltonian 
\begin{equation}
H_{1}(t)=H_{0}+\kappa \mathbf{R}\otimes \delta (t)\mathbf{P}.  \label{1.1}
\end{equation}%
Here $H_{0}=H\otimes {\mathbf{1}}$, where ${\mathbf{1}}$ is the identity
operator in $L^{2}(\Lambda )$, and $\mathbf{R}\otimes \mathbf{P}\equiv \sum
P_{\alpha }R^{\alpha }$, where $P_{\alpha }$, $\alpha =1,\ldots ,d$ are
operator components of the meter momentum vector $\mathbf{P}=(P_{1},\ldots
,P_{d})$ with the pointer coordinate vector $\mathbf{q}=\{q^{\alpha }:\alpha
=1,\ldots ,d\}$ given by the multiplication operators $(q^{\alpha }f)(%
\mathbf{y})=y^{\alpha }f(\mathbf{y})$, $\mathbf{y}\in \Lambda $ in the
Hilbert space $L^{2}(\Lambda )$ of its quantum states such that 
\begin{equation*}
\Vert \mathbf{q}f\Vert ^{2}=\int \sum_{\alpha =1}^{d}|y^{\alpha }f(y)|^{2}%
\mathrm{d}\lambda <\infty \,.
\end{equation*}%
The operators $P_{\alpha }=-\mathrm{i}\hbar \partial /\partial y^{\alpha }$
have the bounded spectrum $(\pi \hbar )[-1/\varepsilon ,1/\varepsilon )^{d}$
for $\Lambda =\varepsilon \mathbb{Z}^{d}$ and are defined by the matrix
elements 
\begin{equation*}
(\mathbf{y}|P_{\alpha }|\mathbf{q})=\frac{1}{(2\pi \hbar )^{d}}\int
p_{\alpha }e^{\mathrm{i}(\mathbf{y}-\mathbf{q})\mathbf{p}/\hbar }\mathrm{d}%
\mathbf{p}\,,\quad \mathbf{q}\in \Lambda \,,
\end{equation*}%
where $(\mathbf{y}|\mathbf{q})=\delta (\mathbf{y}-\mathbf{q})$, $\mathbf{y}%
\in \Lambda $, are the (generalized) eigenfunctions of the coordinate
operators $q^{\alpha }$ normalized with respect to the measure $\mathrm{d}%
\lambda =\varepsilon ^{d}$ in the discrete case $y^{\alpha }\in \varepsilon 
\mathbb{Z}$. This generates the shift operators (\ref{0.2}) in $%
L^{2}(\Lambda )$ with $\mathbf{x}\in \Lambda $ and the scattering operator 
\begin{equation}
S_{t}=\exp \biggl(-\frac{\mathrm{i}}{\hbar }\kappa \mathbf{R}\otimes \mathbf{%
P}_{t}\biggr)=%
\begin{cases}
S\,, & t>0 \\ 
I\,, & t\leqslant 0%
\end{cases}
\label{1.2}
\end{equation}%
in $\mathcal{H}_{1}$, where $\mathbf{P}_{t}=1_{t}\mathbf{P}$, $1_{t}=1$ if $%
t>0$, and $1_{t}=0$ otherwise.

The singular time dependence of the Hamiltonian (\ref{1.1}) makes it
impossible to define the Schr\"{o}dinger equation $\mathrm{i}\hbar \mathrm{d}%
\psi /\mathrm{d}t=H_{1}(t)\psi (t)$ in the usual sense. But one can define a
unitary stochastic evolution $U_{1}(t):\mathcal{H}_{1}\rightarrow \mathcal{H}%
_{1}$ for some $t_{0}<0$ as the single-jump unitary process 
\begin{equation*}
U_{1}(t)=\exp \biggl(-\frac{\mathrm{i}}{\hbar }\int_{t_{0}}^{t}H_{1}(s)%
\mathrm{d}s\biggr)=e^{\mathrm{i}H_{0}(t_{0}-t)/\hbar }S_{t}
\end{equation*}%
provided that $[R^{a},H]=0$. Usually the commutator is not zero, and the
unitary evolution corresponding to (\ref{1.1}) must be redefined in terms of
the solution $\psi (t)=U_{1}(t)\psi _{0}$ to the regularized wave equation.
This can be done in terms of the forward differentials 
\begin{equation*}
\mathrm{d}\psi (t)=\psi (t+\mathrm{d}t)-\psi (t),\quad \mathrm{d}1_{t}=1_{t+%
\mathrm{d}t}-1_{t}\,;
\end{equation*}%
namely, we define the generalized Schr\"{o}dinger equation as 
\begin{equation}
\mathrm{d}\psi (t)+\frac{\mathrm{i}}{\hbar }H_{0}\psi (t)\mathrm{d}%
t=(S-I)\psi (t)\mathrm{d}1_{t}\,,\quad \psi (t_{0})=\psi _{0}\,.  \label{1.3}
\end{equation}

\begin{proposition}
For every $\psi _{0}\in \mathcal{H}$ the difference equation (\ref{1.3}) has
a unique solution corresponding to the initial value $t_{0}\leqslant 0$;
this solution is given by the unitary operator 
\begin{equation*}
U_{1}(t)=U_{0}(t-t_{0})S_{t}(-t_{0}),\,\,\text{ where }\,\,U_{0}(t)=\exp \{-%
\mathrm{i}H_{0}t/\hbar \},\,\,S_{t}(r)=U_{0}^{\dagger }(r)S_{t}U_{0}(r).
\end{equation*}
\end{proposition}

\begin{proof}
Let us rewrite the generalized Schr\"{o}dinger equation in the integral form 
\begin{equation}
\psi (t)=e^{-\mathrm{i}H_{0}t/\hbar }\biggl(e^{\mathrm{i}H_{0}t_{0}/\hbar
}\psi _{0}+\int_{t_{0}}^{t}e^{\mathrm{i}H_{0}r/\hbar }(S-I)\psi (r)\mathrm{d}%
1_{r}\biggr)  \label{1.4}
\end{equation}%
(the equivalence of (\ref{1.4}) to (\ref{1.3}) can be shown by
straightforward differentiation). We can write $\psi (0)=U_{0}(-t_{0})\psi
_{0}$ for the solution (\ref{1.4}) when $t=0$ since $\int_{t_{0}}^{t}\psi (r)%
\mathrm{d}1_{r}=0$ for any $t_{0}\leqslant 0$; hence $\psi (t)$ can be
rewritten as 
\begin{equation*}
\psi (t)=U_{0}(t)(\psi (0)+(S-I)1_{t}\psi (0))=U_{0}(t)S_{t}\psi (0)
\end{equation*}%
for any $t_{0}<0$ since 
\begin{equation*}
\int_{t_{0}}^{t}\psi (r)\mathrm{d}1_{r}=1_{t}\psi (0)
\end{equation*}%
and $(S-I)1_{t}=S_{t}-I$ by the definition of the scattering operator (\ref%
{1.2}). Thus, $U(t)$ is the unitary operator 
\begin{equation*}
U_{0}(t)S_{t}U_{0}(-t_{0})=U_{0}(t-t_{0})S_{t}(-t_{0}).
\end{equation*}%
This gives the solution to equation (\ref{1.3}) for $t_{0}=0$ as well, since 
$\psi (t)=U_{0}(t)S_{t}\psi (0)$ is equal to $\psi _{0}$ for $t=0$.
\end{proof}

The rescaled pointer $\mathbf{q}_{t}=1_{t}\mathbf{q}$ switched on at the
instant $t=0$ of the scattering is described in the space $\mathcal{H}%
\otimes L^{2}(\Lambda )$ by the operators $\mathbf{Q}_{t}=\kappa
^{-1}I\otimes \mathbf{q}_{t}$. In the Heisenberg picture $U_{1}^{\dagger }(t)%
\mathbf{Q}_{t}U_{1}(t)$ these operators are described for any $t_{0}<0$ by
the operators $\mathbf{Y}_{t}(r)=S^{\dagger }(r)\mathbf{Q}_{t}S(r)$ taken at 
$r=-t_{0}$; for all $t>0$ these operators have the shifted form 
\begin{equation}
\mathbf{Y}_{t}(r)=\mathbf{R}(r)\otimes {\mathbf{1}}_{t}+\frac{1}{\kappa }%
\,I\otimes \mathbf{q}_{t}\equiv \mathbf{R}_{t}(r)+\mathbf{Q}_{t}\,,
\label{1.5}
\end{equation}%
where $\mathbf{R}(r)=U^{\dagger }(r)\mathbf{R}U(r)$ and ${\mathbf{1}}_{t}$
is the operator ${\mathbf{1}}$ for $t>0$ and ${\mathbf{0}}$ for $t\leqslant
0 $ in $L^{2}(\Lambda )$. It follows from the commutativity condition 
\begin{equation}
\lbrack Y_{s}^{\alpha },Y_{t}^{\beta }]=0\,,\quad \forall \,s,t;\;\;\alpha
,\beta =1,\ldots ,d  \label{1.6}
\end{equation}%
that the observables $\{\mathbf{Y}_{t}\}$ are selfnondemolition in the sense
of their joint measurability, and they are nondemolition with respect to an
arbitrary particle operator $X_{t}(r)=U_{1}(t)^{\dagger }(X\otimes {\mathbf{1%
}})U_{1}(t)$ at $t_{0}=-r$ in the Heisenberg picture in the sense of their
predictability \cite{9,13} 
\begin{equation*}
\lbrack X_{s},Y_{t}^{\alpha }]=0\,,\quad \forall \,s\geqslant t;\;\;\alpha
=1,\ldots ,d;
\end{equation*}%
indeed, $[X_{s},Y_{s}^{\alpha }]=0$ and $Y_{s}=Y_{t}$ for $s,\,t>0$ since 
\begin{equation*}
U_{1}^{\dagger }(s)\mathbf{Q}_{t}U_{1}(s)=S^{\dagger }(-t_{0})\mathbf{Q}%
_{t}S(-t_{0})\,,\quad \forall \,s\geqslant t\,.
\end{equation*}

Let us fix a state vector $f_{0}\in L^{2}(\Lambda )$, $\Vert f_{0}\Vert =1$,
given by a localized wave function $f_{0}(\mathbf{y})$ on $\Lambda $ at $%
\mathbf{y}=0$. Let $|\mathbf{y})$ denote the (generalized) eigenfunction in
the spectral representation $\mathbf{q}=\int \mathbf{y}|\mathbf{y})(\mathbf{y%
}|\mathrm{d}\lambda $. In the discrete case one can take the sharply
localized $f_{0}=\varepsilon ^{d/2}|0)$ given by the function $f_{0}(\mathbf{%
y})=e(\mathbf{y})/\varepsilon ^{d/2}$, where $e(\mathbf{y})=1$ if $\mathbf{y}%
=0$ and $e(\mathbf{y})=0$ if $\mathbf{y}\neq 0$. This yields the localizing
transformations $F_{t}(\mathbf{y})=(\mathbf{y}|S_{t}f_{0}$ in the form 
\begin{equation*}
F_{t}(\mathbf{y})=f_{0}(\mathbf{y}I-\mathbf{R}_{t})=%
\begin{cases}
f_{0}(\mathbf{y}I-\mathbf{R})\,, & t>0 \\ 
f_{0}(\mathbf{y})I\,, & t\leqslant 0.%
\end{cases}%
\end{equation*}

The reduced transformations $\eta \mapsto \psi (t,\mathbf{y})$, $\mathbf{y}%
\in \Lambda $, defined on the particle space $\mathcal{H}$ by the formula 
\begin{equation*}
\psi (t,\mathbf{y})=U(t-t_{0})F_{t}(-t_{0},\mathbf{y})\eta \,,\quad \mathbf{y%
}\in \Lambda ,
\end{equation*}%
with $U(t)=\exp \{-\mathrm{i}/\hbar H_{t}\}$ and $F_{t}(r)=U^{\dagger
}(r)F_{t}U(r)$ reproduce the unitary evolution $U_{1}(t)$ on $\eta \otimes
f_{0}\in \mathcal{H}_{1}$ similarly to Eq.~(\ref{0.3}) and (\ref{0.4}),
namely, 
\begin{equation*}
\psi (t,\mathbf{y})=U(t)(\mathbf{y}|S_{t}f_{0}U(-t_{0})\eta =(\mathbf{y}%
|U_{1}(t)(\eta \otimes f_{0})\,,\quad \forall \,\eta \in \mathcal{H}\,.
\end{equation*}

The operators $F_{t}(r)$ at $r=-t_{0}$ with an initial wave function $\eta $
of the particle before the scattering ($t_{0}\leqslant 0$) define the
probability measure 
\begin{equation*}
\mu _{t}(\Delta )=\int_{\Delta }\Vert F(t,\mathbf{y})\eta \Vert ^{2}\mathrm{d%
}\lambda =\langle \eta ,\Pi _{t}(\Delta )\eta \rangle \,,\quad \Delta
\subseteq \Lambda ,
\end{equation*}%
for the statistics of the nondemolition measurement of $\kappa \mathbf{R}$
via the observation of the pointer position $\mathbf{y}\in \Delta $ after
the scattering. It is given by a positive operator-valued measure $\Pi
_{t}(\Delta )=\int_{\Delta }|f_{0}(\mathbf{y}-\kappa \mathbf{R}%
_{t}(-t_{0}))|^{2}\mathrm{d}\lambda $, which is normalized, $\Pi
_{t}(\Lambda )=I$, since the measure $\mathrm{d}\lambda $ on $\varepsilon
\Lambda =\mathbb{Z}^{d}$ or $\Lambda =\mathbb{R}^{d}$ is
translation-invariant.

By renormalizing the operators $F(\mathbf{q})$ as $E(\mathbf{y})=\varepsilon
^{d/2}F(\mathbf{y})$ with step $\varepsilon =\kappa $, one obtains the
orthogonal projections 
\begin{equation*}
E(\mathbf{y})=e(\kappa \mathbf{R}-\mathbf{y}I)=\int_{\mathbf{x}:\lfloor 
\mathbf{x}/\kappa \rfloor =\mathbf{y}/\kappa }|\mathbf{x}\rangle \langle 
\mathbf{x}|\mathrm{d}\mathbf{x}\,.
\end{equation*}%
in the case $\kappa \mathbf{R}=\int \lambda (\mathbf{x})|\mathbf{x}\rangle
\langle \mathbf{x}|\mathrm{d}\mathbf{x}$ when $\lambda (\mathbf{x})=\kappa
\lfloor \mathbf{x}/\kappa \rfloor $. Thus, $\Pi _{t}(\Delta )$, $\Delta
\subseteq \Lambda $, is the spectral measure $\sum_{y\in \Delta }E(t_{0},%
\mathbf{y})$ of the quantized position $\kappa \mathbf{R}(-t_{0})$ of the
particle for a $t>0$ given by the eigenorthoprojections $E(r,\mathbf{y}%
)=U^{\dagger }(r)E(\mathbf{y})U(r)$ of the operators $\mathbf{R}(r)$,
corresponding to the rescaled pointer integer values $\mathbf{y}/\kappa $.
Thus, the projection reduction postulate has been deduced \TEXTsymbol{>}from
the Hamiltonian interaction (\ref{1.1}) and the nondemolition measurement
for the sharply localized initial state $f_{0}$. But there is no continuous
limit as $\kappa \rightarrow 0$ of such sharp reduction with nontrivial $%
e\neq 0$, since $\Vert e\Vert ^{2}=\int e(\mathbf{y})\mathrm{d}\lambda
=\kappa ^{d}\rightarrow 0$ and the sharp function $e(y)$ disappears as the
element of the meter state space $L^{2}(\Lambda )$.

In the continuous case, one can take an unsharp $f_{0}\in \Lambda
^{2}(\Lambda )$ and renormalize the operators (\ref{1.8}) as $G_{t}(\mathbf{y%
})=f_{0}(\mathbf{y}I-\kappa \mathbf{R}_{t})/f_{0}(\mathbf{y})$ if $f_{0}(%
\mathbf{y})\neq 0$, as in the Gaussian case (\ref{0.5}). They define $G_{t}(%
\mathbf{y})$ as the identity operator for $t\leqslant 0$ or $r>0$, whereas
for $t>0$ and $r\leqslant 0$ the operator 
\begin{equation}
G(\mathbf{y})=\langle \mathbf{y}|Sf_{0}=(\mathbf{y}|G\,,\quad
G=f_{0}^{-1}Sf_{0}\,,  \label{1.7}
\end{equation}%
say, of the Gaussian form 
\begin{equation*}
G(\mathbf{y})=\exp \{\pi \kappa \mathbf{R}(\mathbf{y}I-\frac{1}{2}\,\kappa 
\mathbf{R})\}
\end{equation*}%
given by the generalized eigenfunctions $|\mathbf{y}\rangle =|\mathbf{y}%
)/f_{0}(\mathbf{y})$ for the spectral representation $\mathbf{q}=\int 
\mathbf{y}|\mathbf{y}\rangle \langle \mathbf{y}|\mathrm{d}\mu _{0}$ with
respect to the initial probability measure $\mathrm{d}\mu _{0}=|f_{0}(%
\mathbf{y})|^{2}\mathrm{d}\lambda $ with density $|f_{0}(\mathbf{y}%
)|^{2}=\exp \{-\pi \mathbf{y}^{2}\}$ in the case (\ref{0.5}). The
corresponding propagators 
\begin{equation*}
T(t,\mathbf{y})=U(t-t_{0})G_{t}(-t_{0},\mathbf{y}),\quad \mathbf{y}\in
\Lambda ,\;t_{0}\leqslant 0
\end{equation*}%
define the operator-valued measure $\Pi _{t}(\Delta )$ as 
\begin{equation*}
\Pi _{t}(\Delta )=\int_{\Delta }T^{\dagger }(t,\mathbf{y})T(t,\mathbf{y})%
\mathrm{d}\mu _{0}=\int_{\Delta }|G_{t}(-t_{0},\mathbf{y})|^{2}\mathrm{d}\mu
_{0},
\end{equation*}%
so that the output probability measure $\mu _{t}(\Delta )=\langle \eta ,\Pi
_{t}(\Delta )\eta \rangle $ is absolutely continuous with respect to $\mu
_{0},\;\mu _{0}(\Delta )=0\Rightarrow \mu _{t}(\Delta )=0$. Hence, the
reduced state vector $\chi (t,\mathbf{y})=T(t,\mathbf{y})\eta $ is
normalized to $1$ as a stochastic vector process $\chi (t):\mathbf{y}\mapsto
\chi (t,\mathbf{y})\in \mathcal{H}$ in the mean square sense with respect to
the input probability measure $\mu _{0}$, 
\begin{equation*}
\Vert \chi (t)\Vert _{0}^{2}=\int \langle \chi (t,\mathbf{y}),\chi (t,%
\mathbf{y})\rangle \mathrm{d}\mu _{0}=\langle \eta ,\eta \rangle =1\,.
\end{equation*}%
This model of nondemolition observation with continuous data $\mathbf{y}\in 
\mathbb{R}^{d}$ also applies to unsharp measurement of operators $\mathbf{R}$
with discrete spectrum. In contrast to sharp measurement, unsharp
measurement is not sensitive to the continuous spectrum limit as $\kappa
\rightarrow 0$ of $\mathbf{R}=\lfloor \mathbf{x}/\kappa \rfloor $,
corresponding to the replacement of $\kappa \mathbf{R}$ by $\mathbf{x}\in 
\mathbb{R}^{d}$.

\begin{theorem}
For any initial $t_{0}\leqslant 0$ the stochastic vector process $\chi
(t)=T(t)\eta $ satisfies the single-kick equation 
\begin{equation*}
\mathrm{d}\chi (t)+\frac{\mathrm{i}}{\hbar }H\chi (t)\mathrm{d}t=\mathrm{d}%
1_{t}[G-I]\chi (t)\,,\quad \chi (t_{0})=\eta ,
\end{equation*}%
generated by the random differential $\mathrm{d}1_{t}[G-I](\mathbf{y})=(G(%
\mathbf{y})-I)\mathrm{d}1_{t}$ on $\mathbf{y}\in \Lambda $ with respect to
the initial probability measure $\mu _{0}$. This simplest reduction equation
is written in terms of the forward differentials $\mathrm{d}\chi (t,\mathbf{y%
})=\chi (t+\mathrm{d}t,\mathbf{y})-\chi (t,\mathbf{y})$, that is, is
understood in the sense of It\^{o}.
\end{theorem}

\begin{proof}
Indeed, by representing $G_{t}$ as $G_{t}(\mathbf{y})=(G(\mathbf{y}%
)-I)1_{t}+I=I+1_{t}[G-I](\mathbf{y})$, for $\chi
(t)=U(t-t_{0})G_{t}(-t_{0})\eta $ one obtains the integral equation 
\begin{align*}
\chi (t)& =U(t-t_{0})\eta +U(t)1_{t}[G-I]U(-t_{0})\eta = \\
& =e^{-\mathrm{i}Ht/\hbar }\biggl(e^{\mathrm{i}Ht_{0}/\hbar }\eta
+\int_{t_{0}}^{t}e^{\mathrm{i}Hr/\hbar }\mathrm{d}1_{r}[G-I]\chi (r)\biggr)%
\,.
\end{align*}%
But this equation is equivalent to the differential equation (\ref{1.10}), a
fact that can be proved by straightforward differentiation taking into
account the It\^{o} multiplication table 
\begin{equation*}
(\mathrm{d}t)^{2}=0\,,\quad \mathrm{d}t\mathrm{d}1_{t}=0=\mathrm{d}1_{t}%
\mathrm{d}t\,,\quad (\mathrm{d}1_{t})^{2}=\mathrm{d}1_{t}\,.
\end{equation*}
\end{proof}

Similarly, one can obtain the simplest nonlinear stochastic equation for the
normalized reduced state vector $\chi _{y}(t)=\chi (t,\mathbf{y})/\Vert \chi
(t,\mathbf{y})\Vert $: 
\begin{equation*}
\mathrm{d}\chi _{y}(t)+\mathrm{i}/\hbar H\chi _{y}(t)\mathrm{d}%
t=(G_{y}(t)-I)\chi _{y}(t)\mathrm{d}1_{t}\,,\quad \chi _{y}(0)=\eta ,
\end{equation*}%
where $G_{y}(t)=G(\mathbf{y})/\Vert G(\mathbf{y})\chi _{y}(t)\Vert $, $\eta
\in \mathcal{H}$. This equation is an equivalent differential form of the
nonlinear integral stochastic equation 
\begin{multline*}
\chi _{y}(t)=e^{-\mathrm{i}Ht/\hbar }\biggl(e^{\mathrm{i}Ht_{0}/\hbar }\eta
+\int_{t_{0}}^{t}e^{\mathrm{i}Hr/\hbar }(G_{y}(r)-I)\chi _{y}(r)\mathrm{d}%
1_{r}\biggr)= \\
=U(t-t_{0})\eta +U(t)(G_{y}-I)1_{t}\chi _{y}(0)=U(t)G_{t}(\mathbf{y})\chi
_{y}(0)/\Vert G_{t}(\mathbf{y})\chi _{y}(0)\Vert .
\end{multline*}%
This yields $\chi _{y}(t)=T(t,\mathbf{y})\eta /\Vert T(t,\mathbf{y})\eta
\Vert $ for an $t_{0}\leqslant 0$ because of $\Vert G_{t}(\mathbf{y})\eta
\Vert =\Vert T(t,\mathbf{y})\eta \Vert $.

Note that the random state vector $\chi _{y}(t)$ is obtained by conditioning
with respect to the output (rather than input) probability measure $\mathrm{d%
}\mu =\Vert \chi (t,\mathbf{y})\Vert ^{2}\mathrm{d}\mu _{0}$.

\section{Spontaneous Localization of a Single Particle}

Let us consider a spontaneous process of scattering interactions (\ref{1.1})
of a quantum particle at random time instants $t_{n}>0$, $t_{1}<t_{2}<\dots $%
, with a renewable meter in an apparatus of the cloud chamber type with
bubbles serving as the meter. We consider the increasing sequences $%
(t_{1},t_{2},\ldots )$ as countable subsets $\tau \subset \mathbb{R}_{+}$
such that $\tau _{t}=\tau \cap \lbrack 0,t)$ is finite for any $t\geqslant 0$
in accordance with the finiteness of the number of scattered bubbles on the
finite observation interval $[0,t)$. The set of all such infinite $\tau $
will be denoted by $\Gamma _{\infty }$, $\Gamma $ is the inductive limit $%
\cup \Gamma _{t}$ as $t\rightarrow \infty $ of $\Gamma _{t}=\{\tau _{t}:\tau
\in \Gamma _{\infty }\}$, which is equal to the disjoint union $\Gamma
_{t}=\sum_{n=0}^{\infty }\Gamma _{t}(n)$ of $n$-simplices $\Gamma
_{t}(n)=\{t_{1}<\dots <t_{n}\}\subset \lbrack 0,t)^{n}$.

The measurement apparatus is assumed to be a quantum system of infinitely
many bubbles each of which is identical to the single meter described in the
previous section. The coordinates $q_{n}^{\alpha }$, $\alpha =1,\dots ,d$,
of a bubble labeled by the scattering number $n\in \mathbb{N}$ show the
position $\mathbf{y}_{n}\in \Lambda $ of the pointer at time $t_{n}\in \tau $%
.

The interaction Hamiltonian of the particle corresponding to these
scatterings is given by the series 
\begin{equation}
H(t,\tau )=H_{0}+\kappa \mathbf{R}\otimes \sum_{n=1}^{\infty }\delta
(t-t_{n})\mathbf{P}(n)  \label{2.1}
\end{equation}%
having at most two nonzero terms when $t\in \tau $. Here $H_{0}=H\otimes {%
\mathbf{1}}$ is the Hamiltonian describing the time evolution on the
intervals between the scatterings $t\in \tau $ and $\mathbf{P}(n)$ is the
momentum of the $n$th scattered bubble, given as the vector-operator $%
\mathbf{P}(n)=(P_{1}(n),\ldots ,P_{d}(n))$, where $P_{\alpha }(n)=-\mathrm{i}%
\hbar \mathrm{d}/\mathrm{d}q_{n}^{\alpha }$ in the case $\Lambda =\mathbb{R}%
^{d}$, and $\mathbf{R}\otimes \mathbf{P}(n)=\sum P_{a}(n)R^{a}$.

The generalized Schr\"{o}dinger equation corresponding to the Hamiltonian (%
\ref{1.1}) can be written for fixed $\tau \in \Gamma _{\infty }$ by analogy
with the single-kick case 
\begin{equation}
\mathrm{d}\psi (t)+\frac{\mathrm{i}}{\hbar }\,H_{0}\psi (t)\,\mathrm{d}%
t=(S(n_{t})-I)\psi (t)\,\mathrm{d}n_{t},\quad \psi (0,\tau )=\psi _{0}.
\label{2.2}
\end{equation}%
Here $S(n)=\exp \{-(\mathrm{i}/\hbar )\kappa \mathbf{R}\otimes \mathbf{P}%
(n)\}$ and $n_{t}(t)=|\tau _{t}|$ is the numerical process that gives the
cardinality $|\tau _{t}|=\sum_{r\in \tau }1_{t-r}$ of the localized subset $%
\tau _{t}=\{t_{n}<t\}$, so that $\mathrm{d}n_{t}(t)$ is equal to $1$ for $%
t\in \tau $, and zero otherwise.

\begin{proposition}
The solution to the equation (\ref{2.2}) is uniquely determined for every $%
\tau \in \Gamma _{\infty }$ by the initial state $\psi _{0}$ of the system.
Namely, $\psi (t,\tau )=U(t,\tau )\psi _{0}$, where $U(t,\tau
)=U_{0}(t)V_{t}^{\dagger }(\tau )$, $V_{t}^{\dagger }(\tau )$ is the
chronological product $\prod_{r\in \tau }^{\leftarrow
}S_{t}(r)=S(t_{n_{t}})\dots S(t_{1})$, and 
\begin{equation}
V_{t}(\tau )=S_{t}^{\dagger }(t_{1})S_{t}^{\dagger }(t_{2})\dots =\biggl(%
\prod_{r\in \tau }^{\leftarrow }S_{t}(r)\biggr)^{\dagger }\,.  \label{2.3}
\end{equation}%
Here $S_{t}(t_{n})=U_{0}^{\dagger }(t_{n})S_{t}(n)U_{0}(t_{n})$ for $t_{n}<t$%
, where $S(n)=\exp \{-(\mathrm{i}/\hbar )\,\kappa \mathbf{R}\otimes \mathbf{P%
}(n)\}$, and $S_{t}(t_{n})=I$ for $t_{n}\geqslant t$, so that the infinite
product (\ref{2.3}) contains only a finite number $n_{t}=\sum_{r\in \tau
}1_{t-r}$ of factors different from the identity operator $I$.
\end{proposition}

\begin{proof}
Recall that the differential equation (\ref{2.2}) is equivalent to the
integral equation given by the recurrence relation 
\begin{equation}
\psi (t,\tau )=e^{-\mathrm{i}H_{0}t/\hbar }\biggl(\psi _{0}+\sum_{r\in \tau
}^{r<t}e^{\mathrm{i}H_{0}r/\hbar }(S(n_{r})-I)\psi (r,\tau )\biggr)
\label{2.4}
\end{equation}%
for every $\tau \in \Gamma _{\infty }$. Hence, $\psi (t,\tau
)=U_{0}(t)V_{t}^{\dagger }(\tau )\psi _{0}$, where $U_{0}(t)=e^{-\mathrm{i}%
H_{0}t/\hbar }$ and $V_{t}(\tau )$ is a solution to the operator equation 
\begin{equation*}
V_{t}(\tau )=I+\sum_{r\in \tau }^{r<t}V_{r}(t)(S_{t}^{\dagger }(r,\tau
)-I)\,,\quad V_{0}(\tau )=I\,,
\end{equation*}%
where $S^{\dagger }(t,\tau )=U_{0}(t)^{\dagger }S(n_{t}(\tau ))^{\dagger
}U_{0}(t)$. But this equation has a unique solution (\ref{2.2}), which can
be written as the binomial sum 
\begin{equation*}
\lbrack L_{t}(t_{1})+I][L_{t}(t_{2})+I]\dots =\sum_{\rho \subseteq \tau
_{t}}L(r_{1},\tau )\dots L(r_{n},\tau )
\end{equation*}%
in terms of $\rho =\{r_{1},\dots ,r_{n}\}$, $r_{1}<\dots <r_{n}$, $%
n\leqslant n_{t}$, $L_{t}(r)=S_{t}^{\dagger }(r)-I$ ($=0$ if $r>t$) and $%
L(r,\tau )=S^{\dagger }(r,\tau )-I$. Indeed, this sum contains $I$ as the
null product corresponding to $r=0$ and the sum of the other terms is equal
to 
\begin{align*}
V_{t}(\tau )-I& =\sum_{r\in \tau }^{r<t}\sum_{\rho \subseteq \tau
_{r}}^{\rho <r}L(r_{1},\tau )\dots L(r_{m},\tau )L(r,\tau ) \\
& =\sum_{r\in \tau }^{r<t}V_{r}(\tau )L(r,\tau )=\sum_{r\in \tau
}^{r<t}V_{r}(\tau )(S^{\dagger }(r,\tau )-I),
\end{align*}%
where $m\leqslant n_{t}-1$.
\end{proof}

Note that the differential equation (\ref{2.2}) depending on $\tau \in
\Gamma _{\infty }$ via $n_{t}=n_{t}(\tau )$ is not stochastic as long as we
have not fixed a probability distribution for the instants $\tau
=(t_{1},t_{2},\dots )$ of the spontaneous interactions. To obtain a
continuous (at least in the mean) dynamics for such an instantaneous
process, one can assume that the probability distribution of the random
number process $n_{t}(\tau )$ is given by the Poisson law $\pi _{0}(\mathrm{d%
}\tau )$ on $\Gamma _{\infty }$ presented as the projective limit as $%
t\rightarrow \infty $ of the probability measures 
\begin{equation}
\pi _{0}(\mathrm{d}\tau _{t})=e^{-\nu t}\nu ^{|\tau _{t}|}\,\mathrm{d}\tau
_{t}\,,\quad \nu >0\,.  \label{2.5}
\end{equation}%
Here $\tau _{t}=\tau $ is a finite time-ordered sequence $\tau
(n)=(t_{1},\dots ,t_{n})\in \Gamma _{t}$ with $n=n_{t}$, $\mathrm{d}\tau
_{t}=\prod_{k=1}^{n_{t}}\,\mathrm{d}t_{k}$ is the measure on $\Gamma _{t}$
given by the sum of product measures $\mathrm{d}t_{1},\dots ,\mathrm{d}t_{n}=%
\mathrm{d}\tau (n)$ on the simplices $\Gamma _{t}(n)$, $\mathrm{d}\tau (0)=1$
on $\Gamma _{t}(0)=\{\varnothing \}$ such that 
\begin{equation*}
\int_{\Gamma _{t}}\nu ^{|\tau |}\mathrm{d}\tau \!:\,=\sum_{n=0}^{\infty }\nu
^{n}\int \dots \int_{0\leqslant t_{1}<\dots <t_{n}<t}\mathrm{d}t_{1}\dots 
\mathrm{d}t_{n}=e^{\nu t}\,.
\end{equation*}%
Note that any other numerical process can be described by a positive density
function $f(\tau )$ with respect to the Poissonian measure, that is, has the
form $f(\tau )\pi (\mathrm{d}\tau )$.

Let us fix an initial state $\varphi _{0}=f_{0}^{\infty }$ of the apparatus
as the infinite product $f_{0}^{\infty }=\otimes _{k=1}^{\infty }f_{k}$ of
the identical state vectors $f_{k}=f_{0}$ of the bubbles given by a
normalized element $f_{0}\in L^{2}(\Lambda )$. We suppose, as in \S 1, that $%
f_{0}(\mathbf{y})\neq 0$ for almost all $\mathbf{y}\neq \Lambda $ such that
the space $\mathcal{E}=\{\varphi :\Lambda \rightarrow \mathbb{C}:\Vert
\varphi \Vert _{0}^{2}<\infty \}$ is isomorphic to $L^{2}(\Lambda )$ with
respect to the multiplication $\varphi (\mathbf{y})=f(\mathbf{y})/f_{0}(%
\mathbf{y})$ by $f_{0}^{-1}$ and the scalar product $\Vert \varphi \Vert
_{0}^{2}=\int |\varphi (\mathbf{y})|^{2}\mathrm{d}\mu _{0}=\Vert f\Vert ^{2}$%
. This defines the solutions $\psi (t,\tau )$ of the stochastic equation (%
\ref{2.2}) with the initial data $\psi _{0}=\eta \otimes \varphi _{0}$ given
by state vectors $\eta \in \mathcal{H}$ in the particle space $\mathcal{H}$
as the state vectors in the product space $\mathcal{H}_{\infty }=\mathcal{H}%
\otimes {\mathcal{E}}_{\infty }$, where ${\mathcal{E}}_{\infty
}=\lim_{n\rightarrow \infty }L^{2}(\Lambda ^{n})$ is the Hilbert space
generated by the infinite-product functions $\varphi (\boldsymbol{y})\simeq
\prod_{k=1}^{\infty }f_{k}(\mathbf{y}_{k})$ with $f_{k}=f_{0}$ for almost
all $k$. One can describe ${\mathcal{E}}_{\infty }$ as the space of all
functions $\varphi :\Lambda ^{\infty }\rightarrow \mathbb{C}$ square
integrable with respect to the product measure $\mu _{0}^{\infty }(\mathrm{d}%
\boldsymbol{y})=\mu _{0}(\mathrm{d}\mathbf{y}_{1})\cdot \mu _{0}(\mathrm{d}%
\mathbf{y}_{2})\cdots $ on the space $\Lambda ^{\infty }=\Lambda \times
\Lambda \times \dots $ of sequences $\boldsymbol{y}=(\mathbf{y}_{1},\mathbf{y%
}_{2},\dots )$: $\Vert \varphi \Vert _{0}^{2}=\int |\varphi (\boldsymbol{y}%
)|^{2}\mu _{0}^{\infty }(\mathrm{d}\boldsymbol{y})<\infty )$. The
generalized product vectors $|\boldsymbol{y}\rangle =|\mathbf{y}_{1}\rangle
\otimes |\mathbf{y}_{2}\rangle \otimes \dots $ of this space are defined by
the $\delta $-functions $\langle \mathbf{q}|\mathbf{y}\rangle $ normalized
with respect to $\mu _{0}:\int |\boldsymbol{y}\rangle \langle \boldsymbol{y}%
|\mu _{0}^{\infty }(\mathrm{d}\boldsymbol{y})=1$.

Consider the sequence $\boldsymbol{q}=(\mathbf{q}_{1},\mathbf{q}_{2},\dots )$
of coordinates $q_{n}^{\alpha }$, $\alpha =1,\dots ,d$, of the scattered
bubbles at the time instants $\{t_{1},t_{2},\dots \}$. The commuting
vector-operators $\mathbf{q}_{n}$, $n\in \mathbb{N}$, described in ${%
\mathcal{E}}_{\infty }$ by the multiplications $\mathbf{q}_{n}|\boldsymbol{y}%
\rangle =|\boldsymbol{y}\rangle \mathbf{y}_{n}$, are assumed to be measured
at the random time instants $t_{n}$, $n\in \mathbb{N}$. The point
trajectories of such measurements are given by the sequences $\boldsymbol{y}%
=(y_{1},y_{2},\dots )$ of pairs $y_{n}=(t_{n},\mathbf{y}_{n})$ with $%
t_{1}<t_{2},\dots $ and $\mathbf{y}_{n}\in \Lambda $, identified with
countable subsets $\upsilon =\{y_{1},y_{2},\dots \}\subset \mathbb{R}%
_{+}\times \Lambda $. As elements $\upsilon =(\tau ,\boldsymbol{y})$ of the
Cartesian product $\Upsilon _{\infty }=\Gamma _{\infty }\times \Lambda
^{\infty }$, they have the probability distribution ${\mathrm{P}}_{0}(%
\mathrm{d}\upsilon )=\pi _{0}(\mathrm{d}\tau )\mu _{0}^{\infty }(\mathrm{d}%
\boldsymbol{y})$, where $\Lambda ^{\infty }$ is the space of all sequences $%
\boldsymbol{y}=(\mathbf{y}_{1},\mathbf{y}_{2},\dots )$, $\mathbf{y}_{n}\in
\Lambda $, equipped with the probability product-measure $\mu _{0}^{\infty }(%
\mathrm{d}\boldsymbol{y})$.

The measurement data of the observable process up to a given time instant $%
t>0$ is described by a finite sequence $\upsilon_t=(y_1,\dots,y_n)$ with $%
n=n_t(\upsilon)$ given by the numerical process $n_t(\tau)$ for the
component $\tau$ of $\upsilon$.

Let us introduce the counting distribution $n_{t}(\Delta )=|\upsilon
_{t}\cap (\mathbb{R}_{+}\times \Delta )|$ as the number $n_{t}(\Delta
,\upsilon )$ of scatterings in the time-space region $[0,t)\times \Delta $
and define the counting integral $\int_{0}^{\infty }\int_{\Lambda }L(r,%
\mathbf{y})\mathrm{d}n_{r}(\mathrm{d}\mathbf{y})$ over $y\in \mathbb{R}%
_{+}\times \Lambda $ as the series 
\begin{equation}
n[L](\upsilon )=\sum_{y\in \upsilon }L(y)\,,\quad \forall \,\upsilon \in
\Upsilon _{\infty }\,.  \label{2.6}
\end{equation}%
Having fixed an integer-valued distribution $n_{t}(\Delta )\in \{0,1,\dots
\} $ as a function of $t\geqslant 0$ and of measurable sets $\Delta
\subseteq \Lambda $, one can obtain the corresponding trajectory $\upsilon $
as a sequence of the counts of the jumps of $n_{t}(\Delta )$ in the
time-space $\mathbb{R}_{+}\times \Lambda $.

Given an initial state vector in $\mathcal{H}_{\infty }$ of the form $\psi
_{0}=\eta \otimes \varphi _{0}$ with fixed $\varphi _{0}\simeq f_{0}^{\infty
}$, one can define a nonunitary stochastic evolution $\eta \mapsto
T(t,\upsilon )\eta $ by setting 
\begin{equation*}
T(t,\upsilon )=\langle \boldsymbol{y}|U(t,\tau )\varphi _{0}\,,\quad
\upsilon =(\tau ,\boldsymbol{y}),
\end{equation*}%
which reproduces the unitary evolution $U(t,\tau )=U_{0}(t)V_{t}^{\dagger
}(\tau )$ defined by (\ref{2.2}). This can also be written as $T(t,\upsilon
)=U(t)F_{t}^{\dagger }(\upsilon )$, since $U_{0}(t)=U(t)\otimes {\mathbf{1}}$
commutes with the (generalized) eigen-bras $\langle \boldsymbol{y}|=\langle 
\mathbf{y}_{1},\mathbf{y}_{2},\dots |$ of the bubble coordinates $(\mathbf{q}%
_{1},\mathbf{q}_{2},\dots ):\langle \boldsymbol{y}|U_{0}(t)=U(t)\langle 
\boldsymbol{y}|$. The reduction transformations $F_{t}(\upsilon )$, $%
\upsilon \in \Upsilon _{\infty }$, are given by the chronological products 
\begin{equation}
F_{t}(v)=G_{t}^{\dagger }(y_{1})G_{t}^{\dagger }(y_{2})\dots \equiv
\prod_{y\in v}^{\rightarrow }G_{t}^{\dagger }(y)  \label{2.7}
\end{equation}%
of $G_{t}(t_{n},\mathbf{y}_{n})=U^{\dagger }(t_{n})G(\mathbf{y}_{n})U(t_{n})$
for $t_{n}<t$, where $G(\mathbf{y})=\langle \mathbf{y}|Sf_{0}$, owing to the
product form (\ref{2.3}) of the unitary transformations $V_{t}(\tau )$, $%
\tau \in \Gamma $, and $\langle \boldsymbol{y}|=\otimes _{k=1}^{\infty
}\langle \mathbf{y}_{k}|$, $\varphi _{0}(\boldsymbol{y})\simeq
\prod_{k=1}^{\infty }f_{0}(\mathbf{y}_{k})$, $\langle \boldsymbol{y}|\varphi
_{0}\equiv 1$ for $\boldsymbol{y}\in \Lambda ^{\infty }$. The stochastic
operator (\ref{2.7}) defined by the single-point reductions 
\begin{equation}
G_{t}(r,\mathbf{y})=\langle \mathbf{y}|S_{t}(r)f_{0}=%
\begin{cases}
G(r,\mathbf{y})\,, & r\leqslant t \\ 
I,\quad & r>t%
\end{cases}
\label{2.8}
\end{equation}%
is normalized with respect to the initial probability ${\mathrm{P}}_{0}(%
\mathrm{d}\upsilon )$, $\Pi _{\Upsilon _{\infty }}[I](t)=I$, where 
\begin{equation}
\Pi _{A}[X](t)=\int_{A}T^{\dagger }(t,\upsilon )XT(t,\upsilon ){\mathrm{P}}%
_{0}(\mathrm{d}\upsilon )=\int_{A}F_{t}(v)XF_{t}^{\dagger }(\upsilon ){%
\mathrm{P}}_{0}(\mathrm{d}\upsilon )\,,  \label{2.9}
\end{equation}%
is a continual operational-valued measure \cite{3}--\cite{5} defined on
measurable sets $A\subseteq \Upsilon _{\infty }$ of the point trajectories $%
\upsilon _{t}=\{(r,\mathbf{y})\in \upsilon |r<t\}$ given by the operations $%
\Phi _{t}(\upsilon ):X\mapsto F_{t}(\upsilon )XF_{t}^{\dagger }(\upsilon )$
for particle operators $X:\mathcal{H}\rightarrow \mathcal{H}$.

The positive operator-valued measure $\Pi _{t}(A)=\Pi _{A}[I](t)$ gives the
statistics 
\begin{equation*}
{\mathrm{P}}_{t}(A)=\int_{A}\Vert (\boldsymbol{y}|U(t,\tau )\psi _{0}\Vert
^{2}\mu _{0}^{\otimes }(\mathrm{d}\boldsymbol{y})\pi _{t}(\mathrm{d}\tau )
\end{equation*}%
for the continual observation with respect to an arbitrary initial
wave-function $\psi _{0}=\eta \otimes \varphi _{0}$, $\eta \in \mathcal{H}$
in the form 
\begin{equation*}
{\mathrm{P}}_{t}(A)=\langle \eta ,\Pi _{t}(A)\eta \rangle \,.
\end{equation*}%
The output probability measure ${\mathrm{P}}(A)$, $A\subseteq \Upsilon
_{\infty }$, is defined by the marginales ${\mathrm{P}}_{t}(A)$, $A\subseteq
\Upsilon _{t}$, as $t\rightarrow \infty $.

\begin{theorem}
The reduced wave function $\chi (t,\upsilon )=T(t,\upsilon )\eta $ is
normalized 
\begin{equation*}
\Vert \chi (t)\Vert ^{2}=\int \Vert \chi (t,\upsilon )\Vert ^{2}{\mathrm{P}}%
_{0}(\mathrm{d}\upsilon )=1
\end{equation*}%
as a stochastic vector process $\chi (t):\Upsilon _{\infty }\rightarrow 
\mathcal{H}$ with respect to the initial probability ${\mathrm{P}}_{0}$. It
satisfies the stochastic wave equation 
\begin{equation}
\mathrm{d}\chi (t)+\frac{\mathrm{i}}{\hbar }\,H\chi (t)\,\mathrm{d}t=\mathrm{%
d}n_{t}[G-I]\chi (t)\,,\quad \chi (0)=\eta ,  \label{2.10}
\end{equation}%
expressed in terms of the random differential $\mathrm{d}n_{t}[G-I](\upsilon
)=(G(\mathbf{y}_{n_{t}}(\upsilon ))-1)\mathrm{d}n_{t}(\upsilon )$, $%
n_{t}(\upsilon )=n_{t}(\tau )$ for the point distribution $%
n_{t}[L]=\int_{\Lambda }L(\mathbf{y})n_{t}(\mathrm{d}\mathbf{y})=n[L_{t}]$
over $\mathbf{y}\in \Lambda $ defined as (\ref{2.6}) with $L_{t}(r,\mathbf{y}%
)=1_{t}(r)L(\mathbf{y})$.
\end{theorem}

\begin{proof}
To prove Eq.~(\ref{2.10}), discovered for the first time in \cite{13}, we
rewrite it in the following integral form 
\begin{equation*}
\chi (t)=e^{-\mathrm{i}Ht/\hbar }\biggl(\eta +\int_{0}^{t}\int_{\Lambda }e^{%
\mathrm{i}Hr/\hbar }(G(\mathbf{y})-I)\chi (r)\mathrm{d}n_{r}(\mathrm{d}%
\mathbf{y})\biggr),
\end{equation*}%
given for each $\upsilon \in \Upsilon _{\infty }$ by the finite sum 
\begin{equation*}
n[1_{t}U^{\dagger }(r)(G-I)\chi ](\upsilon )=\sum_{(r,\mathbf{y})\in
\upsilon }^{r<t}U^{\dagger }(r)(G(\mathbf{y})-I)\chi (r)\,.
\end{equation*}%
We express the solution to this equation in the form $\chi (t,\upsilon
)=U(t)F_{t}^{\dagger }(\upsilon )\eta $ via the solution (\ref{2.7}) to the
recursion equation 
\begin{equation*}
F_{t}(\upsilon )=I+\sum_{(r,\mathbf{y})\in \upsilon }^{r<t}F_{r}(\upsilon
)(G(r,\mathbf{y})-I)\,,\quad F_{0}(\upsilon )=I,
\end{equation*}%
with $G(t,\mathbf{y})=U^{\dagger }(t)G(\mathbf{y})U(t)$, as was done for the
unitary case.
\end{proof}

Let us also write on $F$ the nonlinear equation 
\begin{equation*}
\mathrm{d}\chi _{\upsilon }(t)+\frac{\mathrm{i}}{\hbar }\,H\chi _{\upsilon
}(t)\mathrm{d}t=(G_{\upsilon }(t)-I)\chi _{\upsilon }(t)\mathrm{d}%
n_{t}(\upsilon )\,,\quad \chi _{\upsilon }(0)=\eta ,
\end{equation*}%
with $G_{\upsilon }(t)=G(\mathbf{y}_{n_{t}(\upsilon )})/\Vert G(\mathbf{y}%
_{n_{t}(\upsilon )})\chi _{\upsilon }(t)\Vert $. Its solutions define the
normalized reduction $\chi _{\upsilon }(t)=\chi (t,\upsilon )/\Vert \chi
(t,\upsilon )\Vert $ for continual counting measurements as a stochastic
vector process $\chi _{\upsilon }(t)\in \mathcal{H}$ with respect to the
output probability measure ${\mathrm{P}}$ of the point process $t\mapsto
\upsilon _{t}$. This can easily be obtained, as in \cite{14}, by applying
the It\^{o} multiplication table 
\begin{equation*}
(\mathrm{d}t)^{2}=0\,,\quad \mathrm{d}t\,\mathrm{d}n_{t}=0=\mathrm{d}n_{t}%
\mathrm{d}t\,,\quad (\mathrm{d}n_{t})^{2}=\mathrm{d}n_{t}\,.
\end{equation*}

\section{Mixing Reduction for Many Particles}

We now consider $M$ identical particles interacting independently with the
bubbles in accordance with the scattering term in the Hamiltonian (\ref{2.1}%
). The spontaneous process of scatterings is described by the time-ordered
sequences of pairs $(k_{n},t_{n})$, $t_{1}<t_{2}<\dots $, where $k_{n}\in
\{1,\dots ,M\}$ is the number of the particle labeled by the scattering
number $n\in \mathbb{N}$ at time instant $t_{n}>0$. We have excluded the
possibility of two or more scatterings of the bubbles at the same instant of
time, as was done for a single particle in \S 2. The sequence $(k_{1},t_{1})$%
, $(k_{2},t_{2}),\dots $ of the scatterings can be represented by the
occupational subsets $\tau _{k}=\{t_{n}\in \tau :k_{n}=k\}$ of the time set $%
\tau =\{t_{1},t_{2},\dots \}$, which are disjoint, $\tau _{k}\cap \tau
_{l}=\varnothing $ if $k\neq l$, since the scatterings for different
particles are independent. We shall consider the $M$-tuples $\tau _{\bullet
}=(\tau _{1},\dots ,\tau _{M})$ of these countable subsets $\tau _{k}\subset 
\mathbb{R}_{+}$ as elements $\tau _{\bullet }\in \Gamma _{\infty }^{M}$ of
the Cartesian $M$-product $\Gamma _{\infty }$, given by the partition $\tau
=\sqcup \tau _{k}:\,=\cup \tau _{k}$, $\tau _{k}\cap \tau _{l}=\varnothing $
if $k\neq l$ of a $\tau \in \Gamma _{\infty }$.

The interaction Hamiltonian of $M$ particles with independent scatterings
labled by a sequence $\tau _{\bullet }=\{(k_{1},t_{1}),\,(k_{2},t_{2}),\dots
\}$ reads 
\begin{equation}
H(t,\tau _{\bullet })=H_{0}^{M}+\kappa \sum_{n=1}^{\infty }\mathbf{R}%
(k_{n}))\otimes \delta (t-t_{n})\mathbf{P}(n).  \label{3.1}
\end{equation}%
Here $H_{0}^{M}=H^{M}\otimes {\mathbf{1}}$ is the Hamiltonian of the
particles describing the time evolution on the intervals between the
scatterings with the bubbles: 
\begin{equation*}
H^{M}=\sum_{k=1}^{M}H(k)+\sum_{k=1}^{M}\sum_{l>k}^{M}W(k,l)\,,
\end{equation*}%
where $H(k)=I^{\otimes (k-1)}\otimes H\otimes I^{\otimes (M-k)}$ is the
Hamiltonian of the $k$th particle and $W(k,l)$ is the interaction potential
in $\mathcal{H}^{\otimes M}$ of the $k$th and $l$th particle, $1\leqslant
k<l\leqslant M$.

Let $\mathcal{H}^{M}$ denote the $M$-particle Hilbert space, which is an
invariant subspace $\mathcal{H}^{M}\subseteq \mathcal{H}^{\otimes M}$ of
symmetric (bosons) or antisymmetric (fermions) $M$-tensors $\eta ^{M}\in 
\mathcal{H}^{\otimes M}$ generated by the product-vectors $\otimes
_{k=1}^{M}\eta _{k}\in \mathcal{H}^{\otimes M}$ with $\eta _{k}\in \mathcal{H%
}$. The correspondent It\^{o}-Schr\"{o}dinger equation for the stochastic
state vector $\psi ^{M}(t):\tau _{\bullet }\mapsto \psi (t,\tau _{\bullet })$
with values $\psi (t,\tau _{\bullet })\in \mathcal{H}_{\infty }^{M}$ in the
product space $\mathcal{H}_{\infty }^{M}=\mathcal{H}^{M}\otimes \mathcal{E}%
_{\infty }$ of the $M$-particle space $\mathcal{H}^{M}$ by $\mathcal{E}%
_{\infty }=\lim_{n\rightarrow \infty }L^{2}(\Lambda ^{n})$ reads 
\begin{equation}
\mathrm{d}\psi ^{M}(t)+\frac{\mathrm{i}}{\hbar }\,H_{0}^{M}\psi ^{M}(t)%
\mathrm{d}t=(S(k_{t},n_{t})-I)\psi ^{M}(t)\mathrm{d}n_{t}\,.  \label{3.2}
\end{equation}%
Here $S(k,n)=\exp \{-\frac{\mathrm{i}}{\hbar }\,\kappa \mathbf{R}(k)\otimes 
\mathbf{P}(n)\}$, $\mathbf{R}(k)=I^{\otimes (k-1)}\otimes \mathbf{R}\otimes
I^{\otimes (M-k)}$, 
\begin{equation*}
k_{t}(\tau _{\bullet })=\sum_{k=1}^{M}k1_{\tau _{k}}(t)\,,\quad \text{where}%
\;1_{\tau }(t)=%
\begin{cases}
1\,, & t\in \tau \\ 
0\,, & t\notin \tau%
\end{cases}%
\end{equation*}%
is the random number $k_{t}:\Gamma _{\infty }^{M}\rightarrow \{1,\dots ,M\}$%
, labeling a particle by $k$ at any instant $t\in \tau _{k}$ of its
collision with a bubble labeled by $n_{t}(\tau )=\sum_{k=1}^{M}n_{k,t}=|\tau
\cap (0,t)|$, where $n_{k,t}=|\tau _{k}\cap \lbrack 0,t)|$, $\tau =\cup \tau
_{k}$.

\begin{proposition}
The solutions $\psi (t,\tau _{\bullet })=U(t,\tau _{\bullet })\psi _{0}^{M}$%
, $\psi _{0}^{M}\in \mathcal{H}_{\infty }^{M}$, of Eq.~(\ref{3.2}) can be
written as $U(t,\tau _{\bullet })=U_{0}^{M}(t)V_{t}^{\dagger }(\tau
_{\bullet })$ in terms of the finite chronological product 
\begin{equation}
V_{t}(\tau _{\bullet })=S_{t}^{\dagger }(k_{1},t_{1})S_{t}^{\dagger
}(k_{2},t_{2})\dots ,\quad \tau _{\bullet }\in \Gamma _{\infty }^{M}\,,
\label{3.3}
\end{equation}%
where $S_{t}(k,t_{n})=U_{0}^{M\dagger }(t_{n})S(k,n)U_{0}^{M}(t_{n})$ and $%
U_{0}^{M}(t)=\exp \{-\frac{\mathrm{i}}{\hbar }\,H_{0}^{M}(t)\}$.
\end{proposition}

The proof is exactly the same as for the case of a single particle ($M=1$).

Let $\omega =(w_{1},w_{2},\dots )$ denote a chronologically ordered sequence
of triples $w_{n}=(k_{n},t_{n},\mathbf{y}_{n})$ and $\Omega $ the space of
such sequences with $\{t_{1},t_{2},\dots \}\in \Gamma _{\infty }$. Every
sequence $\omega \in \Omega $ can be represented as a pair $\omega =(\tau
_{\bullet },\boldsymbol{y})$, where $\tau _{\bullet }=(\tau _{1},\dots ,\tau
_{M})$ is a partition of the corresponding sequence $\tau
=\{t_{1},t_{2},\dots \}$ and $\boldsymbol{y}=(\mathbf{y}_{1},\mathbf{y}%
_{2},\dots )$, so that $\Omega $ can be identified with the product $\Gamma
_{\infty }^{M}\times \Lambda ^{\infty }$. The space $\Omega $ is equipped
with the probability measure ${\mathrm{P}}_{0}(\mathrm{d}\omega )=\pi _{0}(%
\mathrm{d}\tau _{\bullet })\mu _{0}^{\infty }(\mathrm{d}\boldsymbol{y})$,
where $\pi _{0}(\mathrm{d}\tau _{1},\dots ,\mathrm{d}\tau
_{M})=\prod_{k=1}^{M}\pi _{0}(\mathrm{d}\tau _{k})$ is the product of the
identical Poisson measures (\ref{2.5}), in accordance with the independence
of spontaneous interactions of each particle with the bubbles.

Given an initial state vector $\psi ^{M}=\eta ^{M}\otimes \varphi _{0}$,
where $\varphi _{0}\simeq f_{0}^{\infty }$, one can easily prove that the
nonunitary stochastic evolution 
\begin{equation*}
T(t,\omega )=\langle \boldsymbol{y}|U(t,\tau _{\bullet })\varphi
_{0}\,,\quad \omega =(\tau _{\bullet },\boldsymbol{y})
\end{equation*}%
is also a finite chronological product 
\begin{equation*}
T(t,\omega )=U^{M}(t)F_{t}^{\dagger }(\omega )\,,\quad F_{t}(\omega
):=G_{t}^{\dagger }(w_{1})G_{t}^{\dagger }(w_{2})\dots \,.
\end{equation*}%
Here $G_{t}(k,t_{n},\mathbf{y}):=I$ for $t_{n}>t$ and 
\begin{equation*}
G_{t}(k,t_{n},\mathbf{y}):=U^{M\dagger }(t_{n})G(k,\mathbf{y}%
)U^{M}(t_{n}),\,\,t_{n}\leqslant t,
\end{equation*}%
where $\,U^{M}(t)=\exp \{-\frac{\mathrm{i}}{\hbar }\,H^{M}t\}$, is defined
by the reduced scattering operator 
\begin{equation}
G(k,\mathbf{y})=I^{\otimes (k-1)}\otimes G(\mathbf{y})\otimes I^{\otimes
(M-k)}\,,\quad G(\mathbf{y})=\langle \mathbf{y}|Sf_{0}  \label{3.4}
\end{equation}%
for $f_{0}\in L^{2}(\Lambda )$, $\Vert f_{0}\Vert =1$, applied to the $k$th
particle only in $\mathcal{H}^{M}$. The stochastic operator $T(t)$ defines
the solutions $\chi ^{M}(t,\omega )=T(t,\omega )\eta ^{M}$ to the It\^{o}
differential equation 
\begin{equation*}
\mathrm{d}\chi ^{M}(t)+\frac{\mathrm{i}}{\hbar }H^{M}\chi ^{M}(t)\,\mathrm{d}%
t=\sum_{k=1}^{M}\,\mathrm{d}n_{k,t}[G(k)-I]\chi ^{M}(t)\,,\quad \chi
^{M}(0)=\eta ^{M}
\end{equation*}%
for the stochastic vector states $\chi ^{M}(t):\Omega \rightarrow \mathcal{H}%
^{M}$ of the $M$-particle system, corresponding to an initial $\eta ^{M}\in 
\mathcal{H}^{M}$. The right-hand side of this equation is written as a point
integral $n_{t}[L]=\sum_{k=1}^{M}\int L(k,\mathbf{y})\,\mathrm{d}%
n_{t}=n[L_{t}]$ with respect to the stochastic distribution $n[L](\omega
)=\sum_{w\in \omega }L(w)$ for $L(k,r,\mathbf{y})=1_{t}(r)L(k,\mathbf{y}%
)\equiv L_{t}(w)$. The vector $\chi ^{M}(t,\omega )\in \mathcal{H}^{M}$ as
well as $\chi (t,\upsilon )$ in (\ref{2.10}), is no longer normalized $%
(\Vert \chi ^{M}(t,\omega )\Vert \neq 1)$ for an initial state-vector $\eta
^{M}\in \mathcal{H}^{M}$, $\Vert \eta ^{M}\Vert =1$, but it is normalized
with respect to the probability measure ${\mathrm{P}}_{0}$ on $\Omega $ in
the mean square sense. But, in contrast to $\chi (t,\upsilon )$, the vector $%
\chi ^{M}(t,\omega )$ is not yet the reduced description of the $M$-particle
system under the observation of the scattering process $\upsilon _{t}=\{(r,%
\mathbf{y})\in \upsilon :r\leqslant t\}$, given by the registration of the
pointer positions $\mathbf{y}_{n}\in \Lambda $ at random time instants $%
t_{n} $.

The reduced dynamics corresponding to the observation is described by a
stochastic operational process $X\mapsto \Theta \lbrack X](t)$, 
\begin{equation}
\Theta \lbrack X](t,\upsilon )=\Phi _{t}[U^{M\dagger }(t)XU^{M}(t)](\upsilon
)  \label{3.5}
\end{equation}%
for the $M$-particle operators $X:\mathcal{H}^{M}\rightarrow \mathcal{H}^{M}$
given by the conditional expectation 
\begin{equation}
\Phi _{t}[X](\tau ,\boldsymbol{y})=\frac{1}{M^{|\tau _{t}|}}\,\sum_{\sqcup
\sigma _{k}=\tau _{t}}F_{t}(\sigma _{\bullet },\boldsymbol{y}%
)XF_{t}^{\dagger }(\sigma _{\bullet },\boldsymbol{y})\,  \label{3.6}
\end{equation}%
where the sum is taken over all partitions $\sigma _{\bullet }=(\sigma
_{1},\dots ,\sigma _{M})$ of a finite subset $\tau _{t}=\tau \cap \lbrack
0,t)$. This averaging is due to the impossibility to detect the
individuality of the identical particles producing indistinguishable effects
on the bubbles by measuring the scatterings of the bubbles.

To prove Eq.~(\ref{3.6}), we need to compare the correlations of $%
F_{t}(\omega )XF_{t}^{\dagger }(\omega )$ and of an arbitrary functional $%
g(\upsilon _{t})$ of the observable point process $\upsilon _{t}$ with the
correlations of (\ref{3.6}) and of $g(\upsilon _{t})$. But by applying the
well-known formula \cite{21} one can easily find 
\begin{equation*}
\int_{\Gamma _{t}^{M}}x(\sigma _{1},\dots ,\sigma _{M})\prod_{k=1}^{M}\,%
\mathrm{d}\sigma _{k}=\int_{\Gamma _{t}}\sum_{\sigma _{k}:\sqcup \sigma
_{k}=\sigma }x(\sigma _{1},\dots ,\sigma _{M})\,\mathrm{d}\sigma
\end{equation*}%
for the multiple point integration that these correlations with respect to
the probability measure ${\mathrm{P}}_{0}$ on $\Omega $ given by the Poisson
law (\ref{2.5}) simply coincide: 
\begin{align*}
& \int_{\Gamma _{t}^{M}}\pi _{t}(\mathrm{d}\sigma _{\bullet })\int_{\Lambda
^{\infty }}g(\sqcup _{k=1}^{M}\sigma _{k},\boldsymbol{y})F(\sigma _{\bullet
},\boldsymbol{y})XF_{t}^{\dagger }(\sigma _{\bullet },\boldsymbol{y})\mu
_{0}^{\infty }(\mathrm{d}\boldsymbol{y})= \\
=& \int_{\Gamma _{t}^{M}}\langle g(\sqcup _{k=1}^{M}\sigma _{k}),X(\sigma
_{1},\dots ,\sigma _{M})\rangle _{0}\prod_{k=1}^{M}e^{-\nu t}\nu ^{|\sigma
_{k}|}\mathrm{d}\sigma _{k} \\
=& \int_{\Gamma _{t}}\langle g(\sigma ),\sum_{\sigma _{k}:\sqcup \sigma
_{k}=\sigma }X(\sigma _{1},\dots ,\sigma _{M})\rangle _{0}e^{-M\nu t}\nu
^{|\sigma |}\mathrm{d}\sigma \\
=& \int_{\Gamma _{t}}\pi _{t}^{M}(\mathrm{d}\sigma )\int_{\Lambda ^{\infty
}}g(\sigma ,\boldsymbol{y})\frac{1}{M^{|\sigma |}}\,\sum_{\tau _{k}:\cup
\tau _{k}=\sigma }F(\omega )XF^{\dagger }(\omega )\mu _{0}^{\infty }(\mathrm{%
d}\boldsymbol{y})\,.
\end{align*}%
Here $\langle \cdot ,\cdot \rangle _{0}$ is the abbreviation for the inner
product in $\mathcal{E}$ of the test function $\upsilon \mapsto g(\tau ,%
\boldsymbol{y})$ with fixed $\tau \in \Gamma _{t}$ and the operator function 
$X_{t}(\tau _{\bullet },\boldsymbol{y})=F_{t}(\tau _{\bullet },\boldsymbol{y}%
)XF_{t}^{\dagger }(\tau _{\bullet },\boldsymbol{y})$ with fixed $\tau
_{\bullet }\in \Gamma _{t}^{M}$ and $\tau =\sqcup _{k=1}^{M}\tau _{k}$. The
probability measure 
\begin{equation}
\pi _{0}^{M}(\mathrm{d}\tau _{t})=\sum_{\sqcup \sigma _{k}=\tau
_{t}}\prod_{k=1}^{M}\pi _{0}(\mathrm{d}\sigma _{k})=e^{-M\nu _{t}}|M\nu
|^{|\tau _{t}|}\mathrm{d}\tau _{t}  \label{3.7}
\end{equation}%
on $\Gamma _{t}$ has the intensity $M\nu $. It is induced by the measure $%
\pi _{0}(\mathrm{d}\tau _{\bullet })$ on $\Gamma _{\infty }^{M}$ with
respect to the particle identification map $\tau _{\bullet }\in \Gamma
_{\infty }^{M}\mapsto \tau =\cup _{k=1}\tau _{k}$, defining the observable
data $v=(\tau ,\boldsymbol{y})$ by the stochastic map $\omega =(\tau
_{\bullet },\boldsymbol{y})\mapsto (\tau =\sqcup _{k=1}^{M}\tau _{k},%
\boldsymbol{y})\in \Upsilon _{\infty }$ on $\omega \in \Omega $.

By the coincidence of the correlations proved above, the stochastic operator
(\ref{3.5}) is indeed the conditional expectation of the stochastic
operators $X_t(\omega)$ with respect to the observable process $\tau =
\upsilon_t$.

In contrast to the pure operations $X\mapsto X_t(\omega)$, the reduction
operation $X\mapsto\Phi_t[X](v)$ preserves the symmetry of the $M$-particle
operators $X$ with respect to particle permutation. It is the least mixing
operation which preserves the indistinguishability of the particles with
respect to the observations of the bubble scatterings, corresponding to the
complete nondemolition measurement of the particles.

Indeed, the reduction operation (\ref{3.6}) can be simply written as the
finite iteration 
\begin{gather*}
\frac{1}{M^{n}}\sum_{k_{1},\dots ,k_{n}=1}^{M}G_{t}^{\dagger
}(k_{1},y_{1})\cdots G_{t}^{\dagger }(k_{n},y_{n})XG_{t}(k_{n},y_{n})\cdots
G_{t}(k_{1},y_{1})= \\
=\Psi _{t}[\dots \Psi _{t}[X](y_{1})\dots ](y_{n})=\Phi _{t}[X](y_{1},\dots
,y_{n})
\end{gather*}%
with $n=|\tau _{t}|$ single mixing reductions 
\begin{equation}
\Psi \lbrack X](y)=\frac{1}{M}\sum_{k=1}^{M}G^{\dagger }(k,y)XG(k,y)\,.
\label{3.8}
\end{equation}%
Given as the arithmetric mean value of the permutations for the pure
operations $X\mapsto G_{t}^{\dagger }(k,y)XG_{t}(k,y)$, corresponding to the
identical operators (\ref{3.4}), the reductions $X\mapsto \Psi _{t}[X](y)$
are permutationally symmetric and are not mixing only if the pure operations
do not break this symmetry.

The mixing property of the reduced stochastic dynamics $t\mapsto \rho
\lbrack X](t,v)$ derived above for the corresponding statistical states 
\begin{equation}
\rho ^{M}[X](t,\upsilon )=\langle \eta ^{M},\Theta \lbrack X](t,\upsilon
)\eta ^{M}\rangle =\func{Tr}\{X\rho ^{M}(t,\upsilon )\}  \label{3.9}
\end{equation}%
gives an increase in the entropy 
\begin{equation*}
\sigma ^{M}(t,\upsilon )=-\func{Tr}\{\rho ^{M}(t,\upsilon )\ln \rho
^{M}(t,\upsilon )\}
\end{equation*}%
for an ensemble of identical particles even under the condition of complete
nondemolition observation. According to (\ref{3.9}), the reduced density
operators $\rho ^{M}(t,\upsilon )$ for the system of $M$ identical particles
gives the probability density 
\begin{equation*}
{\mathrm{p}}^{M}(t,\upsilon _{t})=\func{Tr}\{\rho ^{M}(t,\upsilon _{t})\}={%
\mathrm{P}}^{M}(\mathrm{d}\upsilon _{t})/{\mathrm{P}}_{0}^{M}(\mathrm{d}%
\upsilon _{t})
\end{equation*}%
of the output process $\upsilon _{t}$. Here ${\mathrm{P}}_{0}^{M}=\pi
^{M}\otimes \mu _{0}^{\infty }$ is the probability measure on $\Upsilon
_{\infty }=\Gamma _{\infty }\times \Lambda ^{\infty }$ defined by the
Poisson measure (\ref{3.7}). This means that the a posteriori density
operator $\rho ^{M}(t)$ is defined as a stochastic positive trace class
operator normalized in the mean sense 
\begin{equation*}
\Vert \rho ^{M}(t)\Vert _{1}=\int \func{Tr}\{\rho ^{M}(t,\upsilon )\}{%
\mathrm{P}}_{0}^{M}(\mathrm{d}\upsilon )=1\,.
\end{equation*}

\begin{theorem}
The density $\rho ^{M}(t)$ satisfies the stochastic operator equation 
\begin{equation}
\mathrm{d}\rho ^{M}(t)+\frac{\mathrm{i}}{\hbar }\,[H,\rho ^{M}(t)]\mathrm{d}%
t=\mathrm{d}n_{t}\biggl(\frac{1}{M}\,\sum_{k=1}^{M}G(k)\rho
^{M}(t)G^{\dagger }(k)-\rho ^{M}(t)\biggr),  \label{3.10}
\end{equation}%
which has a unique solution for every initial condition $\rho
^{M}(0,\upsilon )=\rho _{0}^{M}$ given by the density operator $\rho
_{0}^{M} $ for the $M$-particle states $\rho _{0}^{M}[X]=\func{Tr}\{X\rho
_{0}^{M}\}$.
\end{theorem}

\begin{proof}
Let us prove Eq.~(\ref{3.10}) written in the equivalent integral form 
\begin{equation*}
\rho ^{M}[X](t)=\rho ^{M}[X(t)](t)+\int_{0}^{t}\int_{\Lambda }\rho ^{M}[\Psi
\lbrack X(t-r)](\mathbf{y})-X(t-r)](r)\mathrm{d}n_{t}(\mathrm{d}\mathbf{y})
\end{equation*}%
for an $\rho _{0}^{M}[X]=\langle \eta ^{M},X\eta ^{M}\rangle $, $\eta
^{M}\in \mathcal{H}^{M}$, where $X(t)=U^{M}(t)^{\dagger }XU^{M}(t)$, 
\begin{equation*}
U^{M}(t)=e^{-\mathrm{i}H^{M}t/\hbar }\,,\quad \Psi \lbrack X](\mathbf{y})=%
\frac{1}{M}\sum_{k=1}^{M}G^{\dagger }(k,\mathbf{y})XG(k,\mathbf{y})\,.
\end{equation*}%
Taking into account the fact that the stochastic integral (\ref{2.6}) in
this equation is simply a finite sum for every $\upsilon \in \Upsilon
_{\infty }$ and $t$, one can write it as a recursive operator equation 
\begin{equation*}
\Phi _{t}[X](\upsilon )=X+\sum_{(r,y)\in \upsilon }^{r<t}\Phi _{r}[\Psi
\lbrack X](r,\mathbf{y})-X](\upsilon )\,,
\end{equation*}%
for a stochastic operation $\Phi _{t}$, defining the solutions to Eq.~(\ref%
{3.10}), as in (\ref{3.9}), in terms of the composition (\ref{3.5}).

But such recursive equation has a unique solution $\Phi _{t}(\upsilon )=\Psi
_{t}(y_{1})\circ \Psi _{t}(y_{2})\circ \dots $, defined for a $\upsilon
=(\tau ,\boldsymbol{y})\in \Upsilon _{\infty }$ as the chronological
composition of the maps $\Psi _{t}(r,\mathbf{y}):X\mapsto \Psi \lbrack X](r,%
\mathbf{y})$ if $r\leqslant t$ and $\Psi _{t}[X](r,\mathbf{y})=X$ if $r>t$.
This solution can be found by the iterations 
\begin{equation*}
\Phi _{t}(\upsilon )-I=\!\!\sum_{(r,\mathbf{y})\in \upsilon }^{r<t}\Phi
_{r}(\upsilon )\circ \Lambda (r,\mathbf{y})=\!\!\sum_{(r,\mathbf{y})\in
\upsilon _{t}}\biggl(\Lambda (r,\mathbf{y})+\sum_{(s,\mathbf{y})\in
v_{r}}\Phi _{s}(\upsilon )\circ \Lambda (s,\mathbf{y})\biggr)\dots \,,
\end{equation*}%
where $\Lambda (y)=\Psi (y)-{\mathrm{I}}$, ${\mathrm{I}}$ is the identical
map $X\mapsto X$, and by the binomial formula $\Psi _{t}(y_{1})\circ \Psi
_{t}(y_{2})\circ \dots =\sum_{\sigma \subseteq \upsilon _{t}}\Lambda
(z_{1})\circ \dots \circ \Lambda (z_{n})$ in terms of $\sigma =\{z_{1},\dots
,z_{n}\}$, $z=(r,\mathbf{y})$, $r_{1}<\dots <r_{n}$, $n\leqslant n_{t}$.
\end{proof}

Let us also write the nonlinear stochastic equation 
\begin{equation*}
\mathrm{d}\rho _{\upsilon }^{M}(t)+\frac{\mathrm{i}}{\hbar }\,[H,\rho
_{\upsilon }^{M}(t)]\,\mathrm{d}t=\rho _{\upsilon }^{M}(t)\circ (\Psi
_{\upsilon }(t)-{\mathrm{I}})\,\mathrm{d}n_{t}(\upsilon )\,,
\end{equation*}%
where $\Psi _{\upsilon }(t)=\Psi (\mathbf{y}_{n_{t}(\upsilon )})/\func{Tr}%
\{E(\mathbf{y}_{n_{t}(\upsilon )})\rho \}$, 
\begin{equation*}
\rho \circ \Psi (\mathbf{y})=\frac{1}{M}\,\sum_{k=1}^{M}G(k,\mathbf{y})\rho
G^{\dagger }(k,\mathbf{y})\,,\quad E(\mathbf{y})=\frac{1}{M}%
\,\sum_{k=1}^{M}G^{\dagger }(k,\mathbf{y})G(k,\mathbf{y}))
\end{equation*}%
for the normalized density operator $\rho _{\upsilon }^{M}(t)=\rho
^{M}(t,\upsilon )/{\mathrm{p}}^{M}(t,\upsilon )$. This describes the
conditional expectations $\rho _{\upsilon }^{M}[X](t)=\func{Tr}\{X\rho
_{\upsilon }^{M}(t)\}$ of the $M$-particle operators with respect to the
output probability measure ${\mathrm{P}}^{M}(\mathrm{d}\upsilon )={\mathrm{p}%
}^{M}(t,\upsilon ){\mathrm{P}}_{0}^{M}(\mathrm{d}\upsilon )$, where ${%
\mathrm{p}}^{M}(t,\upsilon )=\func{Tr}\{\rho ^{M}(t,\upsilon )\}$.

\section{Macroscopic and Central Limits of the Model}

We now consider the mean field approximation of the measurement apparatus
fixing its total effect $\nu \kappa =-\gamma $ given by the mean number $\nu 
$ of scattered bubbles per second and an interaction constant $\kappa $
coupling each bubble to a particle in the Hamiltonian (\ref{2.1}). We look
for the limits of the unitary and reduced evolutions (\ref{2.2}) and (\ref%
{2.10}) as $\nu \rightarrow \infty $ and $\kappa \rightarrow 0$ such that $%
\gamma $ is a real constant. To perform these limit passages, we need the
expansions 
\begin{align}
S(n)& =I\otimes {\mathbf{1}}-\mathrm{i}\frac{\kappa }{\hbar }\,\mathbf{R}%
\otimes \mathbf{P}(n)-\biggl(\frac{1}{2}\,\biggr)\biggl(\frac{\kappa }{\hbar 
}\biggr)^{2}\,(\mathbf{R}\otimes \mathbf{P}(n))^{2}+\ldots  \notag \\
G(\mathbf{y})& =I-\kappa \frac{f_{0}^{\prime }(\mathbf{y})}{f_{0}(\mathbf{y})%
}\,\mathbf{R}+\frac{1}{2}\,\kappa ^{2}\mathbf{R}\frac{f_{0}^{\prime \prime }(%
\mathbf{y})}{f_{0}(\mathbf{y})}\,\mathbf{R}+\ldots  \label{4.1}
\end{align}%
of the scattering operator $S(n)=\exp \{-\frac{\mathrm{i}}{\hbar }\,\kappa 
\mathbf{R}\otimes \mathbf{P}(n)\}$ and the reduced operator $G(\mathbf{y}%
)=f_{0}(\mathbf{y}I-\kappa \mathbf{R})/f_{0}(\mathbf{y})$ with respect to
the coupling constant $\kappa $. The first term of the expansion for $S(n)$
disappears in the right-hand side in Eq.~(\ref{2.2}), whereas the second and
third terms give rise to the differentials of the operator-valued stochastic
integrals 
\begin{equation*}
\hat{n}_{t}[\mathbf{P}]=\int_{0}^{t}\mathbf{P}(n_{r})\,\mathrm{d}%
n_{r}\,,\quad \hat{n}_{t}[\mathbf{P}^{2}]=\int_{0}^{t}\mathbf{P}(n_{r})^{2}\,%
\mathrm{d}n_{r}\,.
\end{equation*}%
The corresponding terms 
\begin{equation*}
n_{t}\biggl(\frac{f_{0}^{\prime }}{f_{0}}\biggr)=\int_{0}^{t}\int_{\Lambda }%
\frac{f_{0}^{\prime }(\mathbf{y})}{f_{0}(\mathbf{y})}\,\mathrm{d}n_{r}(%
\mathrm{d}\mathbf{y})\,,\quad n_{t}\biggl(\frac{f^{\prime \prime }}{f_{0}}%
\biggr)=\int_{0}^{t}\int_{\Lambda }\frac{f_{0}^{\prime \prime }(\mathbf{y})}{%
f_{0}(\mathbf{y})}\,\mathrm{d}n_{r}(\mathrm{d}\mathbf{y})
\end{equation*}%
on the right-hand side in Eq.~(\ref{2.10}) can also be written as the
integrals $\hat{n}_{t}[L]=\int_{0}^{t}L(n_{r})\,\mathrm{d}n_{r}$ with values
in operator functions of $\tau \in \Gamma _{\infty }$ 
\begin{equation}
\hat{n}_{t}[L](\tau )=\sum_{n=1}^{n_{t}(\tau )}L(n)\,,\quad L(n)={\mathbf{1}}%
^{\otimes (n-1)}\otimes L\otimes {\mathbf{1}}^{\otimes \infty }\,,
\label{4.2}
\end{equation}%
where $L$ is one of the multiplication operators 
\begin{equation*}
L^{\prime }=f_{0}^{\prime }(\mathbf{q})/f_{0}(\mathbf{q})\,,\quad L^{\prime
\prime }=f_{0}^{\prime \prime }(\mathbf{q})/f_{0}(\mathbf{q})\,.
\end{equation*}%
Here $f_{0}^{\prime }(\mathbf{q})$ denotes the vector of the derivatives $%
\partial _{\alpha }f_{0}(\mathbf{q})=\partial f_{0}(\mathbf{q})/\partial
q^{\alpha }$ and $f^{\prime \prime }(\mathbf{q})$ denotes the matrix with
elements $\partial _{\alpha }\partial _{\beta }f(\mathbf{q})$. The operator $%
\hat{n}_{t}[L](\tau )$ corresponding to a function $L=l(\mathbf{q})$ of the
bubble coordinate vector-operator $\mathbf{q}$ in $L^{2}(\Lambda )$ acts in $%
\mathcal{E}_{\infty }$ as the multiplication operator 
\begin{equation*}
\lbrack \hat{n}_{t}[L](\tau )\varphi ](\boldsymbol{y})=\sum_{n=1}^{n_{t}(%
\tau )}l(\mathbf{y}_{n})\varphi (\boldsymbol{y})\equiv n_{t}[l](\tau ,%
\mathbf{\upsilon })\varphi (\boldsymbol{y})\,.
\end{equation*}

Hence, the main terms on the right-hand side in Eq.~(\ref{2.2}) and (\ref%
{2.10}) for $\kappa \rightarrow 0$ are given by the renormalized stochastic
integrals 
\begin{equation}
\hat{\lambda}(t)=\frac{1}{\nu t}\,\int_{0}^{t}L(n_{r})\,\mathrm{d}n_{r}=%
\frac{1}{\nu t}\,\hat{n}_{t}[L]  \label{4.3}
\end{equation}%
of the operator-valued stochastic functions $L(t,\tau )=L(n_{t}(\tau ))$
with respect to the numerical process $n_{t}(\tau )$ that has the Poisson
probability distribution (\ref{2.5}) on $\Gamma _{\infty }$.

To pass to the large number limit $\nu \rightarrow \infty $ in (\ref{4.4})
for an arbitrary operator $L$ in $L^{2}(\Lambda )$, we need to use the
quantum stochastic representation \cite{22} of the integral (\ref{4.4}) in
the Fock space $\mathcal{F}$ over $L^{2}(\mathbb{R}_{+}\times \Lambda )$.
The space $\mathcal{F}$ is defined as the $L^{2}(\Upsilon )$-space of all
square integrable functions $\phi :\Upsilon \rightarrow \mathbb{C}$, $\Vert
\phi \Vert ^{2}=\int_{\Upsilon }|\phi (\upsilon )|^{2}\lambda (\mathrm{d}%
\upsilon )<\infty $ of time ordered finite sequences $\upsilon
=(y_{1},\ldots ,y_{n})$, $y=(t,\mathbf{y})$ identified with subsets $%
\upsilon \subset \mathbb{R}_{+}\times \Lambda $ of finite cardinality $%
|\upsilon |=0,1,2,\dots $. The measure $\lambda (\mathrm{d}\upsilon )$ on
the union 
\begin{equation*}
\Upsilon =\sum_{n=0}^{\infty }\Upsilon (n)
\end{equation*}%
of the disjoint subsets $\Upsilon (n)=\{\upsilon \in \Upsilon :|\upsilon
|=n\}$ is given as the sum 
\begin{equation*}
\lambda (A)=\sum_{n=0}^{\infty }\lambda (\Upsilon (n)\cap A)
\end{equation*}%
of the product $\lambda (\mathrm{d}\upsilon )=\prod_{y\in \upsilon }\mathrm{d%
}y$ of measures $\mathrm{d}y=\mathrm{d}t\,\mathrm{d}\lambda $ on $\mathbb{R}%
_{+}\times \Lambda $ such that 
\begin{equation*}
\Vert f\Vert ^{2}=\sum_{n=0}^{\infty }\int \!\!\int_{0\leq t_{1}<\ldots \leq
t_{n}<\infty }|f(y_{1},\ldots ,y_{n})|^{2}\prod_{i=1}^{n}\,\mathrm{d}y_{i}\,.
\end{equation*}%
Let $N_{t}[L]$ denote the numerical integral as an operator 
\begin{equation}
(N_{t}[L]\phi )(\tau ,\boldsymbol{y})=\sum_{n=1}^{n_{t}(\tau )}L(n)\varphi
(\tau ,\boldsymbol{y})=\hat{n}_{t}[L](\tau )\phi (\tau ,\boldsymbol{y})
\label{4.4}
\end{equation}%
in the Fock space $\phi \in L^{2}(\Upsilon )$. This integral represents the
operator-valued stochastic integral (\ref{4.2}) by pointwise multiplication
of the functions $\phi (\upsilon )=\phi (t,\boldsymbol{y})$ of $\upsilon \in
\Upsilon $, by $\hat{n}_{t}[L](\tau )$, which is considered as the function
of a finite sequence $\tau \in \Gamma $ because of its independence of $\tau
_{\lbrack t}=\{t_{n}\geq t\}$. In order to obtain the initial probability
measure ${\mathrm{P}}_{0}(\mathrm{d}\upsilon )=\pi (\mathrm{d}\tau )\mu
_{0}^{\infty }(\mathrm{d}\boldsymbol{y})$ on $\Upsilon _{\infty }$ induced
by an initial Fock vector $\phi _{0}\in L^{2}(\Upsilon )$, we need an
isomorphic transformation of (\ref{4.5}) 
\begin{equation}
\hat{N}_{t}[L]=N_{t}[L]+\sqrt{\nu }(A_{t}[f_{0}^{\dagger }L]+A_{t}^{\dagger
}[Lf_{0}])+\nu tf_{0}^{\dagger }Lf_{0}  \label{4.5}
\end{equation}%
which can be locally performed by a unitary transformation 
\begin{equation*}
\hat{N}_{t}[L]=U_{s}^{\dagger }N_{t}[L]U_{s}\,,\quad U_{s}=\exp \{\sqrt{\nu }%
(A_{s}^{\dagger }[f_{0}]-(A_{s}[f_{0}^{\dagger }])\}
\end{equation*}%
for every $t<s$. Here $A_{t}^{\dagger }[f]$ and $A_{t}[f^{\dagger }]$ are
the creation and annihilation integrals of $f\in L^{2}(\Lambda )$, $%
f^{\dagger }\in L^{2}(\Lambda )^{\ast }$, given by the operators 
\begin{align*}
\Big(A_{t}^{\dagger }[f]\phi \Big)(\upsilon )& =\sum_{y\in \upsilon _{t}}f(%
\mathbf{y})\phi (\upsilon \backslash y), \\
\Big(A_{t}[f^{\dagger }]\phi \Big)(\upsilon )& =\int_{[0,t)\times \Lambda }f(%
\mathbf{y})^{\ast }\phi (\upsilon \sqcup y)\mathrm{d}y
\end{align*}%
in the Fock space $L^{2}(\Upsilon )$, where $\upsilon \backslash y$ means
the sequence $\upsilon \in \Upsilon $ with deleted $y=(r,\mathbf{y})$, $r<t$%
, and $\upsilon \cup y$ means the sequence $\upsilon \in \Upsilon $ with an
additional element $y\notin \upsilon $. The characteristic functional of the
stochastic operators $\hat{n}_{t}[L]$ with respect to the initial
state-vector $\varphi _{0}\simeq f_{0}^{\infty }$ in ${\mathcal{E}}_{\infty
} $ and the Poisson probability measure (\ref{2.5}) is now given simply by
the vacuum expectation 
\begin{equation*}
\int_{\Gamma ^{\infty }}(\varphi _{0},e^{\mathrm{i}\hat{n}_{t}[L]}\varphi
_{0})\pi _{0}(\mathrm{d}\tau )=\langle \delta _{\phi },e^{\mathrm{i}\hat{N}%
_{t}[L]}\delta _{\phi }\rangle \,,
\end{equation*}%
where $\delta _{\phi }(\upsilon )=1$ if $\upsilon =\emptyset $; otherwise, $%
\delta _{\phi }(\upsilon )=0$.

The corresponding representation $\hat l(t)=\frac1{\nu t}\, \hat N_t[L]$ for
(\ref{4.4}) helps us immediately obtain the quantum large number limit 
\begin{equation*}
\lim_{\nu\to\infty}\frac 1{\nu t}\,\hat N_t[L]=f^\dagger_0Lf_0\hat 1
\end{equation*}
as the mean value $l_0=(f_0,Lf_0)\equiv f^\dagger_0Lf_0$ of a single-bubble
operator with respect to an initial wave packet $f_0\in L^2(\Lambda)$. This
gives the following:

\begin{proposition}
The macroscopic limit 
\begin{equation*}
\mathrm{d}\psi (t)+\frac{\mathrm{i}}{\hbar }\,H_{0}\psi (t)\,\mathrm{d}t=%
\frac{\mathrm{i}}{\hbar }\,\gamma (\mathbf{R}\otimes \mathbf{p}_{0}\hat{1}%
)\psi (t)\,\mathrm{d}t
\end{equation*}%
of the generalized Schr\"{o}dinger equation (\ref{2.2}) turns out to be
nonsingular with an additional potential $-\gamma \mathbf{p}_{0}\mathbf{R}$
corresponding to the mean momentum $\mathbf{p}_{0}=(f_{0},\mathbf{P}f_{0})$
of a bubble in the initial state $f_{0}$. It coincides with the large number
limit $\nu \rightarrow \infty $ of the reduction equation (\ref{2.10}) under 
$\kappa =-\gamma /\nu $.
\end{proposition}

\begin{proof}
Indeed, the mean field dynamics preserves the product structure $\psi
(t,\upsilon )=\eta (t)\phi _{0}(\upsilon )$ of the initial product-vector $%
\psi _{0}=\eta \otimes \phi _{0}$ on the Fock component $\phi _{0}\in 
\mathcal{F}$ because it is identical by virtue of $H_{0}=H\otimes \hat{1}$.
But unexpectedly (compare with \cite{19}) the large number limit 
\begin{equation}
\mathrm{d}\chi (t)+\frac{\mathrm{i}}{\hbar }\,H\chi (t)\,\mathrm{d}t=\frac{%
\mathrm{i}}{\hbar }\,\gamma \mathbf{p}_{0}\mathbf{R}\chi (t)\,\mathrm{d}t
\label{4.6}
\end{equation}%
of the stochastic nonunitary equation (\ref{2.10}) corresponds to the same
unitary dynamics $\eta (t)=U(t)\eta =\chi (t)$ of the particle state vector
if $\chi (0)=\eta $, since 
\begin{equation*}
\frac{1}{\nu t}\,n_{t}\biggl(\frac{f_{0}^{\prime }}{f_{0}}\biggr)%
\,\rightarrow (f_{0},L^{\prime }f_{0})=\int f_{0}(\mathbf{y})^{\ast
}f_{0}^{\prime }(\mathbf{y})\,\mathrm{d}\mathbf{y}=\frac{\mathrm{i}}{\hbar }%
\,\mathbf{p}_{0}\,.
\end{equation*}
\end{proof}

Note that the macroscopic limits for the $M$-particle system also give
essentially the same type of unitary evolutions in the large space $\mathcal{%
H}^M\otimes\mathcal{F}$ and in the reduced space $\mathcal{H}^M$. To get
this correspondence, one has only to replace the measure (\ref{2.5}) on $%
\Gamma_\infty$ for Eq.~(\ref{3.2}~) by the product measure $\pi^{\otimes
M}_0 $ on $\Gamma^M_\infty$ and for Eq.~(\ref{3.10}) by the induced measure (%
\ref{3.7}) on $\Gamma_\infty$, taking $M\nu$ instead of $\nu$ in (\ref{4.5}%
). This means that the mixing property of the reduced equation (\ref{3.10})
vanishes in the mean field approximation for the bubble system.

Let us now pay attention to the fluctuations with respect to the obtained
large number limits. Such fluctuations might appear for $-\kappa =\gamma
/\nu \rightarrow 0$ in the large time scale $t\sim 1/\kappa $. We can get
these fluctuations without rescaling the time $t$ if we assume that $p_{0}=0$
and $-\kappa =\gamma /\sqrt{\nu }$, so that we have to take into account
also the $\kappa ^{2}$-terms in (\ref{4.1}).

It follows from the Fock space representation (\ref{4.6}) that the quantum
central limit 
\begin{equation*}
\lim_{\nu \rightarrow \infty }\frac{l}{\sqrt{\nu }}\,\hat{N}%
_{t}[L]=A_{t}[f_{0}^{\dagger }L]+A_{t}^{\dagger }[Lf_{0}]
\end{equation*}%
exists for any single-bubble operator $L$ with zero mean value $%
(f_{0},Lf_{0})=0$. We first apply this central limit theorem to the
right-hand side in (\ref{2.2}) represented in $\mathcal{H}\otimes \mathcal{F}
$ as 
\begin{equation*}
\hat{n}_{t}[S-I]=\mathrm{i}{\frac{\gamma }{\hbar }}{\frac{1}{\sqrt{\nu }}}%
\hat{N}_{t}[\mathbf{R}\otimes \mathbf{P}]-{\frac{1}{2}}\biggl({\frac{\gamma 
}{\hbar }}\biggr){\frac{1}{\nu }}N_{t}[(\mathbf{R}\otimes \mathbf{P}%
)^{2}]+\ldots \;.
\end{equation*}%
This yields $\lim \frac{1}{\sqrt{\nu }}\,\hat{N}_{t}[\mathbf{P}]=\hat{%
\mathbf{u}}_{t}$, $\lim \frac{1}{\nu }\,\hat{N}_{t}[P_{\alpha }P_{\beta
}]=(P_{\alpha }f_{0},P_{\beta }f_{0})t$, and 
\begin{equation}
\mathrm{d}\psi (t)+K_{0}\psi (t)\,\mathrm{d}t=\frac{\mathrm{i}}{\hbar }%
\,\gamma (\mathbf{R}\otimes \,\mathrm{d}\hat{\mathbf{u}}_{t})\psi (t)\,.
\label{4.7}
\end{equation}%
Here $K_{0}=K\otimes \hat{1}$, $K=\frac{\mathrm{i}}{\hbar }\,H+\frac{1}{2}%
\biggl(\frac{\gamma }{\hbar }\biggr)^{2}\mathbf{R}\mathbf{\sigma }^{2}%
\mathbf{R}$, $\mathbf{\sigma }^{2}=[\sigma _{\alpha \beta }^{2}]$, 
\begin{equation*}
\hat{\mathbf{u}}_{t}=A_{t}[f_{0}^{\dagger }\mathbf{P}]+A_{t}^{\dagger }[%
\mathbf{P}f_{0}]=2\Re A_{t}^{\dagger }[\mathbf{P}f_{0}]
\end{equation*}%
is a Fock space representation of the Wiener vector process $\boldsymbol{u}%
:t\mapsto \mathbf{u}_{t}$ with the correlations $\sigma _{\alpha \beta
}^{2}=\hbar ^{2}(\partial _{\alpha }f_{0},\partial _{\beta }f_{0})$, $\alpha
,\beta =1,\ldots ,d$, defined by the momentum operators $\mathbf{P}$ due to
the quantum stochastic multiplication formula \cite{21, 22} 
\begin{equation*}
\mathrm{d}\hat{u}_{\alpha }\,\mathrm{d}\hat{u}_{\beta }=\mathrm{d}%
A_{t}[f_{0}^{\dagger }P_{\alpha }]\,\mathrm{d}A_{t}^{\dagger }[P_{\beta
}f_{0}]=f_{0}^{\dagger }P_{\alpha }P_{\beta }f_{0}\,\mathrm{d}t\,.
\end{equation*}%
The central limit equation (\ref{4.8}) for the unitary evolution of the
coupled system turns out to be a stochastic Schr\"{o}dinger-It\^{o} equation
of diffusive type driven by the Wiener process $\boldsymbol{u}=\left\{ 
\mathbf{u}_{t}:t\in \mathbb{R}_{+}\right\} $. The same conclusion obviously
holds for the $M$-particle system driven by $M$ independent Wiener processes 
$\hat{\mathbf{u}}_{t}(k)$ identical to $\hat{\mathbf{u}}_{t}$, $k=1,\ldots
,M $: 
\begin{equation*}
\mathrm{d}\psi ^{M}(t)+K_{0}^{M}\psi (t)\,\mathrm{d}t=\frac{\mathrm{i}}{%
\hbar }\,\gamma \biggl(\sum_{k=1}^{M}\mathbf{R}(k)\otimes \,\mathrm{d}\hat{%
\mathbf{u}}_{t}(k)\biggr)\psi ^{M}(t)\,,
\end{equation*}%
where $K^{M}=\frac{\mathrm{i}}{\hbar }\,H^{M}+\frac{1}{2}\biggl(\frac{\gamma 
}{\hbar }\,\biggr)^{2}\sum_{k=1}^{M}\mathbf{R}(k)\mathbf{\sigma }^{2}\mathbf{%
R}(k)$, $K_{0}^{M}=K^{M}\otimes \hat{1}$.

We now prove that the application of the central limit theorem to the
reduction equation (\ref{2.10}) yields an essentially different type of the
stochastic evolution, originally derived in \cite{21} by quantum calculus
method.

\begin{theorem}
The central limit $\nu =(\gamma /\kappa )^{2}\rightarrow \infty $ of the
stochastic wave equation (\ref{2.10}) has the following nonunitary diffusive
type 
\begin{equation}
\mathrm{d}\chi (t)+K\chi (t)\,\mathrm{d}t=\gamma \mathbf{R}\chi (t)\,\mathrm{%
d}\hat{\mathbf{v}}_{t}\,.  \label{4.8}
\end{equation}%
Here $\mathbf{v}_{t}$ is a complex Wiener vector-process with the Fock-space
representation 
\begin{equation}
\hat{\mathbf{v}}_{t}=A_{t}[f_{0}^{\dagger }L^{\prime }]+A_{t}^{\dagger
}[L^{\prime }f_{0}]=\Re A_{t}^{\dagger }[(\mathbf{w}_{0}+\bar{\mathbf{w}}%
_{0})f_{0}]+\mathrm{i}\Im A_{t}^{\dagger }[(\mathbf{w}_{0}-\bar{\mathbf{w}}%
_{0})f_{0}]  \label{4.9}
\end{equation}%
given by the complex osmotic velocity $\mathbf{w}_{0}(\mathbf{y})=\mathbf{%
\partial }\ln f_{0}(\mathbf{y})$, $\mathbf{\partial }=(\partial _{1},\ldots
,\partial _{d})$ of a single bubble, and the operator $K$ is the same as in (%
\ref{4.8}).
\end{theorem}

\begin{proof}
Indeed, the central limit of the right-hand side in (\ref{2.10}) with $%
\kappa =-\gamma /\sqrt{\nu }$ yields 
\begin{align*}
\lim \mathrm{d}n_{t}[G-I]& =\gamma \mathbf{R}\lim {\frac{1}{\sqrt{\nu }}}\,%
\mathrm{d}n_{t}\biggl({\frac{f_{0}^{\prime }}{f_{0}}}\biggr)+\frac{1}{2}%
\,\gamma ^{2}\mathbf{R}\lim \frac{1}{\nu }\,\mathrm{d}n_{t}\biggl({\frac{%
f_{0}^{\prime \prime }}{f_{0}}}\biggr)\mathbf{R} \\
& =\gamma \mathbf{R}\lim \frac{1}{\sqrt{\nu }}\,\mathrm{d}\hat{N}%
_{t}[L^{\prime }]+\frac{1}{2}\,\mathbf{R}\lim \frac{1}{\nu }\,\mathrm{d}\hat{%
N}_{t}[L^{\prime \prime }]\mathbf{R} \\
& =\gamma \mathbf{R}\,\mathrm{d}\hat{\mathbf{v}}_{t}-\frac{1}{2}\,\biggl(%
\frac{\gamma }{\hbar }\biggr)^{2}\mathbf{R}\mathbf{\sigma }^{2}\mathbf{R}\,%
\mathrm{d}t\,,
\end{align*}%
where the representation (\ref{4.6}) is used for 
\begin{equation*}
\hat{n}_{t}[L^{\prime }]=n_{t}\biggl({\frac{f_{0}^{\prime }}{f_{0}}}\biggr)%
\,\,\text{ and }\,\,\hat{n}_{t}[L^{\prime \prime }]=n_{t}\biggl({\frac{%
f_{0}^{\prime \prime }}{f_{0}}}\biggr)
\end{equation*}%
so that $f_{0}^{\dagger }\partial _{\alpha }\partial _{\beta
}f_{0}=-(\partial _{\alpha }f_{0},\partial _{\beta }f_{0})$ are the matrix
elements of $\lim \frac{1}{\nu t}\,\hat{N}_{t}[L^{\prime \prime
}]=f_{0}^{\dagger }L^{\prime \prime }f_{0}$. The linear stochastic equation (%
\ref{4.9}) obtained has a unique solution $\chi (t,\boldsymbol{v})=T(t,%
\boldsymbol{v})\eta $ for a given $\chi (0,\boldsymbol{v})=\eta \in \mathcal{%
H}$ which is not normalized $\Vert \chi (t,\boldsymbol{v})\Vert \neq 0$ for
every Wiener trajectory $\boldsymbol{v}:t\mapsto \mathbf{v}_{t}$ but is
normalized in the mean square sense $\int \Vert \chi (t,\boldsymbol{v})\Vert
^{2}{\mathrm{P}}_{0}(\mathrm{d}\boldsymbol{v})=1$ with respect to the
Gaussian probability measure ${\mathrm{P}}_{0}$ of $\boldsymbol{v}=\{\mathbf{%
v}_{t}:t\in \mathbb{R}_{+}\}$. The measure ${\mathrm{P}}_{0}$ is defined by
the zero mean values of $\mathbf{v}_{t}$ and by the table 
\begin{align*}
\mathrm{d}\hat{v}_{\alpha }\,\mathrm{d}\hat{v}_{\beta }& =\mathrm{d}%
A_{t}[f_{0}^{\dagger }L_{\alpha }^{\prime }]\mathrm{d}A_{t}^{\dagger
}[L_{\beta }^{\prime }f_{0}]=f_{0}^{\dagger }L_{\alpha }^{\prime }L_{\beta
}^{\prime }f_{0}\,\mathrm{d}t \\
\mathrm{d}\hat{v}_{\alpha }^{\ast }\,\mathrm{d}\hat{v}_{\beta }& =\mathrm{d}%
A_{t}[f_{0}^{\dagger }L_{\alpha }^{\prime \dagger }]\mathrm{d}A_{t}^{\dagger
}[L_{\beta }^{\prime }f_{0}]=f_{0}^{\dagger }\hat{L}_{\alpha }^{\prime
\dagger }L_{\beta }^{\prime }f_{0}\,\mathrm{d}t
\end{align*}%
of commuting multiplications 
\begin{equation*}
\mathrm{d}\hat{v}_{\alpha }\,\mathrm{d}\hat{v}_{\beta }=\mathrm{d}\hat{v}%
_{\beta }\,\mathrm{d}\hat{v}_{\alpha }\,,\quad \mathrm{d}\hat{v}_{\alpha }\,%
\mathrm{d}\hat{v}_{\beta }^{\ast }=\mathrm{d}\hat{v}_{\beta }^{\ast }\,%
\mathrm{d}\hat{v}_{\alpha }\,.
\end{equation*}%
But the reduction noise $\hat{\mathbf{v}}_{t}$ obtained in the Fock space
representation does not commute with the real Wiener process $\hat{\mathbf{u}%
}_{t}=\hat{\mathbf{u}}_{t}^{\ast }$ in (\ref{4.8}), 
\begin{align*}
\mathrm{d}\hat{v}_{\alpha }\,\mathrm{d}\hat{u}_{\beta }& =\mathrm{d}%
A_{t}[f_{0}^{\dagger }]\,\mathrm{d}A_{t}^{\dagger }[P_{\beta
}f_{0}]=f_{0}^{\dagger }L_{\alpha }^{\prime }P_{\beta }f_{0}\,\mathrm{d}t \\
\mathrm{d}\hat{u}_{\alpha }\,\mathrm{d}\hat{v}_{\beta }& =\mathrm{d}%
A_{t}[f_{0}^{\dagger }]\,\mathrm{d}A_{t}^{\dagger }[L_{\beta }^{\prime
}f_{0}]=f_{0}^{\dagger }P_{\alpha }L_{\beta }^{\prime }f_{0}\,\mathrm{d}t
\end{align*}%
if $g_{\alpha \beta }(\mathbf{y})\equiv \partial _{\alpha }\partial _{\beta
}\ln f_{0}(\mathbf{y})\neq 0$, since 
\begin{equation*}
\lbrack \mathrm{d}\hat{v}_{\alpha },\mathrm{d}\hat{u}_{\beta
}]=f_{0}^{\dagger }[L_{\alpha }^{\prime },P_{\beta }]f_{0}\,\mathrm{d}t=%
\frac{\hbar }{\mathrm{i}}\,(f_{0},g_{\alpha \beta }f_{0})\,\mathrm{d}t\,.
\end{equation*}
\end{proof}

In the same way one can obtain the continuous reduction equation 
\begin{multline}
\mathrm{d}\rho ^{M}(t)+(K\rho ^{M}(t)+\rho ^{M}(t)K^{\dagger })\mathrm{d}t=
\\
\biggl(\frac{\gamma }{\hbar }\biggr)^{2}\!\sum_{k=1}^{M}\mathbf{R}(k)\mathbf{%
\sigma }^{2}\rho ^{M}(t)\mathbf{R}(k)\,\mathrm{d}t+\gamma (\mathrm{d}\mathbf{%
w}_{t}\mathbf{R}\rho ^{M}(t)+\rho ^{M}(t)\mathbf{R}\,\mathrm{d}\mathbf{w}%
_{t}^{\ast }),  \label{4.10}
\end{multline}%
$\mathbf{R}=\frac{1}{M}\sum_{k=1}^{M}\mathbf{R}(k)$, where the operator $%
K=K^{M}$ is the same as in the equation for the unitary evolution of the $M$%
-particle system coupled with the bubbles. The derived stochastic equation
for the $M$-particle density operator $\rho ^{M}(t,\boldsymbol{w})$
normalized in the mean is driven by the complex Wiener process $\mathbf{w}%
_{t}=\mathbf{v}_{t}^{M}$ having the same multiplication table as the process 
$\sqrt{M}\mathbf{v}_{t}$ in (\ref{4.9}). The diffusive type equation (\ref%
{4.10}) as (\ref{3.10}) also has the mixing property. This was also derived
in \cite{5} by operational method.


\begin{thebibliography}{10}
\bibitem[1]{1} Ludwig,~G. \emph{Commun. Math. Phys.}, \textbf{9} 1, 1968.

\bibitem[2]{2} Davies,~E.~B. and Lewis,~J. \emph{Commun. Math. Phys}, 
\textbf{17} pp.239--260, 1970.

\bibitem[3]{3} Barchielli,~A., Lanz,~L., and Prosperi, G.~M. \emph{Nuovo
Cimento}, \textbf{72B} 79, 1982.

\bibitem[4]{4} Belavkin, V. P. Theory of control of observable quantum
systems \emph{Automatica and Remote Control}, \textbf{44} 2 pp.178--188,
1983.

\bibitem[5]{5} Belavkin, V.~P. and Barchielli,~A. Measurements continuous in
time and posteriori states in quantum mechanics \emph{J. Phys. A. Math. Gen.}%
, \textbf{24} pp.1495--1514, 1991.

\bibitem[6]{6} J.~von Neumann ``Mathematical Foundations of Quantum
Mechanics``, Princeton University Press, 1955.

\bibitem[7]{7} Peres,~A. \emph{Am.J.Phys.}, \textbf{52} 644 yr 1984.

\bibitem[8]{8} Balentine,~L.~E. \emph{Int.J.Theor.Phys.}, \textbf{27}
pp.211--218, 1987.

\bibitem[9]{9} Belavkin,~V.~P. A new wave equation for a continuous
non-demolition measurement \emph{Phys Letters A}, \textbf{140} 78
pp.355--358, 1989.

\bibitem[10]{10} Belavkin,~V.~P. and Staszewski,~P. A quantum particle
undergoing continuous observation \emph{Phys Letters A}, \textbf{140}
pp.359--362, 1989.

\bibitem[11]{11} Belavkin,~V.~P. A posterior {S}chr\"{o}dinger equation for
continuous non-demolition measurement \emph{J. of Math. Phys.}, \textbf{31}
12 pp.2930--2934, 1990.

\bibitem[12]{12} Belavkin,~V.~P. and Staszewski,~P. Nondemolition
observation of a free quantum particle \emph{Phys. Rev. A.}, \textbf{45} 3
pp.1347--1356, 1992.

\bibitem[13]{13} Belavkin,~V.~P. A continuous counting observation and
posterior quantum dynamics \emph{J. Phys. A Math. Gen.}, \textbf{22} pp. L
1109--L 1114, 1989.

\bibitem[14]{14} Belavkin,~V.~P. A stochastic posterior {S}chr\"{o}dinger
equation for counting non-demolition measurement \emph{Letters in Math. Phys.%
}, \textbf{20} pp.85--89, 1990.

\bibitem[15]{15} Belavkin,~V.~P. and Staszewski,~P. A continuous observation
of photon emission \emph{Rep. on Math. Phys.}, \textbf{29} 2 pp.213--225 ,
1991.

\bibitem[16]{16} Pearle,~P. \emph{Phys.Rev.}, \textbf{D29} 235, 1984.

\bibitem[17]{17} Gisin,~N. \emph{J.Phys.A.}, \textbf{19} pp.205--210, 1986.
Math.Gen.

\bibitem[18]{18} Di\'osi,~L. \emph{Phys.Rev.} \textbf{A40} pp.1165--1174,
1988.

\bibitem[19]{19} Ghirardi,~G.~C., Rimini,~A., and Weber,~T. \emph{Phys. Rev.}%
, \textbf{34D} 470, 1986.

\bibitem[20]{20} Ghirardi,~G.~C., Pearle,~P., and Rimini,~A. Markov
processes in Hilbert space and continuous spontaneous localization of
systems of indentical particles \emph{Phys. Rev. A.}, \textbf{42} pp.78--89,
1990.

\bibitem[21]{21} Belavkin,~V.~P. Chaotic states and stochastic integrations
in quantum systems \emph{Uspekhi Mat. Nauk} \textbf{1}, 1992.

\bibitem[22]{22} Hudson,~R.~L. and Parthasarathy,~K.~R. Quantum It\^o's
formula and stochastic evolution \emph{Comm. Math. Phys.}, \textbf{93}
pp.301--323, 1984.
\end{thebibliography}
\end{document}